\documentclass[10pt,twocolumn,secnumarabic, nobibnotes, aps, prd,superscriptaddress,floatfix]{revtex4-2}
\usepackage{eurosym} 
\usepackage{xcolor,soul}
\usepackage{amsmath}
\usepackage{wrapfig}
\usepackage[utf8]{inputenc}                                          
\usepackage{lipsum} 
\usepackage{subfigure}
\usepackage{wrapfig}
\usepackage{xparse,xcoffins}
\usepackage{calligra}
\usepackage{graphicx}
\usepackage{textcomp}%
\usepackage{braket}
\sethlcolor{red}
\usepackage{amsfonts}

\DeclareMathAlphabet{\mathcalligra}{T1}{calligra}{m}{n}
\DeclareFontShape{T1}{calligra}{m}{n}{<->s*[2.2]callig15}{}

\ExplSyntaxOn
\NewCoffin\imagecoffin
\NewCoffin\labelcoffin
\keys_define:nn { miguel/label }
 {
  label   .tl_set:N = \l_miguel_label_tl,
  labelbox .bool_set:N = \l_miguel_label_box_bool,
  labelbox .default:n = true,
  fontsize .tl_set:N = \l_miguel_label_size_tl,
  fontsize .initial:n = \footnotesize,
  pos .choice:,
  pos/nw .code:n = \tl_set:Nn \l_miguel_label_pos_tl { left,up },
  pos/ne .code:n = \tl_set:Nn \l_miguel_label_pos_tl { right,up },
  pos/sw .code:n = \tl_set:Nn \l_miguel_label_pos_tl { left,down },
  pos/se .code:n = \tl_set:Nn \l_miguel_label_pos_tl { right,down },
  pos/n .code:n = \tl_set:Nn \l_miguel_label_pos_tl { hc,up },
  pos/w .code:n = \tl_set:Nn \l_miguel_label_pos_tl { left,vc },
  pos/s .code:n = \tl_set:Nn \l_miguel_label_pos_tl { hc,down },
  pos/e .code:n = \tl_set:Nn \l_miguel_label_pos_tl { right,vc },
  pos .initial:n = nw,
  xpos .tl_set:N = \l_miguel_label_x_tl,
  ypos .tl_set:N = \l_miguel_label_y_tl,
  xpos .initial:n = ,
  ypos .initial:n = ,
  unknown .code:n   = \clist_put_right:Nx \l_miguel_label_clist
                       { \l_keys_key_tl = \exp_not:n { #1 } }
 }
\clist_new:N \l_miguel_label_clist
\box_new:N \l_miguel_label_box
\box_new:N \l_miguel_label_image_box
\NewDocumentCommand{\xincludegraphics}{O{}m}
 {
  \group_begin:
  \tl_clear:N \l_miguel_label_tl
  \clist_clear:N \l_miguel_label_clist
  \keys_set:nn { miguel/label } { #1 }
  \tl_if_empty:NTF \l_miguel_label_tl
   {
    \miguel_includegraphics:Vn \l_miguel_label_clist { #2 }
   }
   {
    \SetHorizontalCoffin\imagecoffin
     {
      \miguel_includegraphics:Vn \l_miguel_label_clist { #2 }
     }
    \SetHorizontalCoffin\labelcoffin
     {
      \raisebox{\depth}
       {
        \bool_if:NTF \l_miguel_label_box_bool
         { \fcolorbox{white}{white}{\l_miguel_label_size_tl\l_miguel_label_tl} }
         { \l_miguel_label_size_tl\l_miguel_label_tl }
       }
     }
    \SetVerticalPole\imagecoffin{left}{3pt+\CoffinWidth\labelcoffin/2}
    \SetVerticalPole\imagecoffin{right}{\Width-3pt-\CoffinWidth\labelcoffin/2}
    \SetHorizontalPole\imagecoffin{up}{\Height-3pt-\CoffinHeight\labelcoffin/2}
    \SetHorizontalPole\imagecoffin{down}{3pt+\CoffinHeight\labelcoffin/2}
    \use:x{\JoinCoffins\imagecoffin[\l_miguel_label_pos_tl]\labelcoffin[vc,hc]}
    \TypesetCoffin\imagecoffin
   }
   \group_end:
 }
\NewDocumentCommand{\setlabel}{m}
 {
  \keys_set:nn { miguel/label } { #1 }
 }
\cs_new_protected:Nn \miguel_includegraphics:nn
 {
  \includegraphics[#1]{#2}
 }
\cs_generate_variant:Nn \miguel_includegraphics:nn { V }
\ExplSyntaxOff

\makeatletter
\newcommand{\twocolumncaption}{\@dblarg\@twocolumncaption}
\def\@twocolumncaption[#1]#2{  \renewcommand{\@makecaption}[2]{    \par\vskip\abovecaptionskip\begingroup\small\rmfamily
    \splittopskip=0pt
    \setbox\@tempboxa=\vbox{
      \@arrayparboxrestore \let \\\@normalcr
      \hsize=.5\hsize \advance\hsize-1em
      \let\\\heading@cr
      \noindent ##1\ ##2\par    }    \vbadness=10000
    \setbox\z@=\vsplit\@tempboxa to .55\ht\@tempboxa
    \setbox\z@=\vtop{\hrule height 0pt \unvbox\z@}
    \setbox\tw@=\vtop{\hrule height 0pt \unvbox\@tempboxa}
    \noindent\box\z@\hfill\box\tw@\par
    \endgroup\vskip \belowcaptionskip
  }  \setlength{\abovecaptionskip}{4ex}  \caption[#1]{#2}}

\begin{document}

\title{Characterizing Noise Effects on Multipartite Entanglement via Phase-Space Visualization}
\author{B Nithya Priya}
\email{b\textunderscore nithyapriya@cb.students.amrita.edu}
\author{S. Saravana Veni}
\email{s\textunderscore saravanaveni@cb.amrita.edu.in}
\affiliation{Department of Physics, Amrita School of Physical Sciences,
Amrita Vishwa Vidyapeetham, Coimbatore, 641112, India. }
\author{Araceli Venegas-Gomez}
\email{araceli.venegas-gomez@qureca.com}
\affiliation{QURECA Ltd., Glasgow, G2 4JR, Scotland, United Kingdom.}
\author{Ria Rushin Joseph}
\email{ria.joseph@deakin.edu.au}
\affiliation{School of IT,Deakin University,Melbourne, Australia}

\begin{abstract}
This paper investigates the behavior of two fundamental types of multipartite entangled states, namely GHZ(3) and W(3) states under Gaussian-distributed amplitude perturbations and White noise model. The Uhlmann-Jozsa fidelity is taken to be the quantitative measure to show the overall degradation of the quantum states, and is implemented via TQIX : a tool specifically designed for quantum state measurement and  related applications. While fidelity analysis captures the progressive decay of quantum states under noise, it offers only limited understanding regarding the state decay and doesn't provide a detailed analysis of how entanglement structures respond to noise models. To reveal the phase-space characteristics and nonclassical signatures of three-qubit entangled states, we employ the spin Wigner function using equal-angle projection. This approach reveals a continuous fading of quantum coherence with increasing noise strength, ultimately providing a clear picture of transition toward classical-like behavior in phase space. This combined qualitative-quantitative framework provides deeper understanding of how different entanglement structures respond to noise, offering practical applications for designing and implementing noise resilient protocols in quantum computing, and quantum information processing. 
\end{abstract}

\maketitle


\section{Introduction}
Quantum entanglement  is one of the most fundamental principles of quantum mechanics , as it reveals a unique property of correlations among the particles \cite{horodecki2009quantum}. The phenomenon where the state of no single particle in a group can be described independently of the state of other particles, even at larger distances, is termed "quantum entanglement." As a result, measuring the physical properties such as position, momentum, and polarization of one particle exhibits strong correlations with that of the other particle \cite{kocher1967polarization} even if they are light years apart \cite{Yin_2013}. This property of entanglement serves as the primary feature of quantum mechanics, which cannot be justified using the laws of Newtonian mechanics. These nonclassical correlations were first discussed by Albert Einstein , Boris Podolsky, and Nathan Rosen  in 1935 and later termed the EPR paradox \cite{Schrödinger_1935, Schrödinger_1936, Yin_2013}, which questioned the completeness of quantum mechanics  based on the local realism principle \cite{Bancal_2012}. Einstein and his collaborators argued that no physical influence can travel faster than light \cite{roger2004road} and suggested the Local Hidden Variable theory \cite{freedman1972experimental}. Further experiments of Bell inequalities  confirmed the predictions of quantum mechanics , thereby violating local realism \cite{hensen2015loophole, freedman1972experimental}. These results demonstrated that quantum correlations caused by entanglement are not due to ignorance but an intrinsic property of quantum mechanics. Greenberger, Horne, and Zeilinger, beyond Bell inequalities \cite{greenberger1989going}, showed that the entangled states with three or more particles directly contradict the hidden variable theory without the need of inequalities \cite{greenberger1989going, greenberger1990bell, mermin1990quantum}. This is known as multipartite entanglement \cite{Ma2024}, where the states share correlations between three or more particles; unlike bipartite entangled systems, they offer more complex correlations. Quantum entanglement, being the heart of quantum mechanics, has a broad application in the field of quantum information theory and quantum computing, which majorly includes superdense coding, quantum teleportation \cite{matson2012quantum}, and some protocols of quantum cryptography \cite{article}, especially quantum key distributions (QKD) \cite{Renner_2005, BBM92, PRXQuantum_CKA}.

However, in realistic physical implementations, quantum systems are unavoidably exposed to environmental interactions  that introduce noise and decoherence, leading to degradation of entanglement and loss of information \cite{NielsenChuang, Yu2009, Breuer2002}. Hence, understanding how different classes of entangled states respond to noise and perturbations is essential for evaluating their suitability and reliability in practical scenarios \cite{Ma2024, Universe2019}. Our study focuses on two most common, yet fundamentally different, three-qubit entangled states, namely Greenberger, Horne, and Zeilinger (GHZ(3)) and W(3) states \cite{greenberger1989going, Duer2000} under Gaussian-distrubuted amplitude perturbations and White noise model. Fidelity and equal-angle spin Wigner function are used as qualitative and quantitative tools to explore how quantum states respond to noise models \cite{Jozsa1994, Chaturvedi2006}.

The numerical analysis presented in this work was carried out using TQIX: a Toolbox for Quantum in X: quantum Measurement, quantum tomography, quantum metrology, and others \cite{tuan2021tqix}. This toolbox provides efficient functions for generating multipartite quantum states and implementing noise models.The study was performed for three-qubit GHZ and W states , whose mathematical definitions and theoretical properties are discussed in the upcoming sections. All the simulations were performed with a fixed random seed to ensure reproducibility. Gaussian perturbations was introduced to the state vector level by directly perturbing the amplitudes of the quantum state with zero mean and varying standard deviation ($\sigma$) \cite{tuan2021tqix}. Both single-realization and ensemble-averaged scenarios were considered in order to show the difference between instantaneous noise effects from averaged decoherence behavior. White noise was implemented at the density matrix level using the depolarizing channel provided by TQIX \cite{tuan2021tqix}. Since this model intrinsically represents an ensemble-averaged process, no additional averaging was required.The robustness of GHZ(3) and W(3) states under noise was quantitatively analyzed using Uhlmann-Jozsa fidelity. Fidelity was evaluated as a function of noise strength for both Gaussian and White noise. To quantitatively examine the effect of noise on multipartite entanglement and non-classical features, equal-angle spin Wigner functions were employed \cite{Chaturvedi2006}. The Wigner function was evaluated on a grid of polar and azimuthal angles and visualized using two-dimensional contour plots and three-dimensional surface representations. As stated above, for gaussian-distributed amplitude perturbation, Wigner functions were computed for both single realizations and ensemble-averaged density matrices .

The paper is formatted as follows: Section II presents the theoretical background for the study that includes mathematical definitions and properties of GHZ(3), W(3) states, and the classical noise models, Ullmann-Jozsa fidelity, density matrix formalism, and purity, followed by the equal-angle spin Wigner function used for phase-space representation. Section III demonstrates the numerical results, providing a comparative analysis of qualitative and quantitative metrics under different noise models. Finally, section IV concludes the paper with the summary of key findings and outline of directions for future research.

 \section{Quantum Noise Models and Phase-Space Formalism for Multipartite States}
 When more than two particles are involved, entanglement exists in different strengths, which can be categorized as entangled only between some particles and entangled among all the particles simultaneously, leading to biseparable and genuine multipartite entanglement, respectively \cite{articl}. Mathematically speaking, for a mixed state with density matrix $\rho_k$,

\[
 \rho_k = \sum_i p_i \ket{\psi^i_k} \bra{\psi^i_k}
\]
 is completely separable if it can be decomposed into k-pure states \cite{articl, gabriel2010criterionkseparabilitymixedmultipartite}. As stated before, our study examines the effects of Gaussian-distributed amplitude perturbations and white noise on two majorly used tripartite entangled states GHZ(3) and W(3).
\subsection{Inequivalent Classes of Tripartite Entanglement: GHZ and W}
The GHZ(n) state involves at least three subsystems (qubits) and is named after the three authors who first formulated the state—Greenberger, Horne, and Zeilinger in 1989 \cite{greenberger1989going, greenberger1990bell}. Mathematically,

\begin{equation}
\ket{GHZ}_n = \frac{1}{\sqrt{2}}(\ket{0}^{\otimes n} + \ket{1}^{\otimes n}).
\end{equation}

which says that all the n qubits will either be in the $\ket{0}$ state or the $\ket{1}$ state. GHZ states can be viewed as the generalization or extension of Bell states from bipartite entanglement to multipartite entangled states or systems. GHZ states exhibit strong global correlations, and hence they are called maximally entangled multipartite states \cite{Duer2000}. There are two important significances of GHZ states: (i) they provide an all-versus-nothing demonstration of quantum non-locality, accounting for a stronger contradiction to local hidden variable theory compared to Bell states, and (ii) they play a crucial role in various quantum information protocols, including quantum secret sharing and distributed quantum computation \cite{Hillery1999}. Despite their strong non-classical correlations, GHZ states are known to be highly sensitive to environmental noise, making them particularly suitable for studying decoherence effects in multipartite systems.  

The W(n) state can also be generalized to n qubits, defined as an equal quantum superposition of all possible pure states with exactly one particle (qubit) being in the excited state ($\ket{1}$) and the others being in the ground state ($\ket{0}$). Mathematically,

\begin{equation}
\ket{W}_n= \frac{1}{\sqrt{n}} (\ket{100...0} +\ket{0100...0} + \ket{00...1})
\end{equation}

The state is named after Wolfgang Dür, Guifré Vidal, and Ignacio Cirac, who first described the state together in 2000 \cite{Duer2000}. 

Our present work examines the effect of classical noise on tripartite GHZ and W states specifically, mathematically given as

\begin{equation}
\ket{GHZ}_3 = \frac{1}{\sqrt{2}} (\ket{000} + \ket{111})
\end{equation}

\begin{equation}
\ket{W}_3 = \frac{1}{\sqrt{3}} (\ket{100} + \ket{010} + \ket{001})
\end{equation}

For two qubits, entanglement is described as either separable or entangled. But this is not the case in three or more qubits. There are inequivalent kinds of entanglement. One kind of entanglement cannot be converted into other kinds using LOCC \cite{Duer2000}. So, GHZ and W are fundamentally different kinds; they are not just stronger/weaker versions of each other.

The GHZ state is defined as maximally entangled because all three qubits are globally entangled, and it strongly violates the theory of local realism.
\[
\rho_{GHZ} =\ket{GHZ}\bra{GHZ}_3
\]
Tracing out or measuring the third qubit gives
\begin{equation}
Tr_3(\rho_{GHZ}) = \frac{1}{2}(\ket{00}\bra{00} + \ket{11}\bra{11})
\end{equation}
which is separable. This physically means, even if one qubit is lost, the entire entanglement is collapsed. Equation (5) is the classical correlation, not quantum correlation. This confirms the global entanglement of the GHZ state as stated before.

On the other hand, tracing out or measuring the third qubit in the W(3) state gives

\begin{align}
\mathrm{Tr}_3(\rho_W)
&= \frac{1}{3}\Big(
\ket{00}\bra{00}
+ \ket{01}\bra{01}
+ \ket{10}\bra{10} \nonumber\\
&\qquad\quad
+ \ket{10}\bra{01}
+ \ket{01}\bra{10}
\Big)
\end{align}

Which is still an inseparably mixed state \cite{Duer2000}. Equation (6) confirms the distributed/local entanglement property of the W state.
\subsection{Modeling Classical Noise: Gaussian and White Noise}
Gaussian and white noise are widely used classical stochastic models for describing randomness and imperfections in physical systems \cite{Gardiner2004,Carvalho2004}. In this work, gaussian-distributed amplitude perturbation is employed in the form of Gaussian-distributed numerical perturbations, rather than as a specific quantum dynamical noise channel. gaussian-distributed amplitude perturbation may be defined as a random signal that follows normal distribution (bell-shaped curve). This is often characterized using quantities like mean, which is mostly zero, and standard deviation and is given by,
\begin{equation}
P(x) = \frac{1}{\sqrt{2\pi\sigma^2}}e^{-\frac{(x - \mu)^2}{2\sigma^2}}
\end{equation}
where $\mu$ represents the mean and $\sigma$ denotes the standard deviation, which characterizes the strength of the noise. In the classical point of view, it can be shortly said that gaussian-distributed amplitude perturbation is the theoretical model of natural disturbances like the thermal noise in systems \cite{Gardiner2004}.

In quantum information studies, amplitude perturbations are often introduced numerically to assess the sensitivity and robustness of quantum states to imperfections and uncertainties. In this work, Gaussian-distributed perturbations are applied at the state-vector level as a numerical model, rather than as a physical quantum noise channel. This procedure consists of adding random Gaussian perturbations $\delta_n$ to each amplitude of the quantum state, followed by renormalization, \begin{equation} \ket{\psi} \longrightarrow \ket{\psi'}=\frac{1}{\mathcal{N}}\sum_n (\ket{\psi_n}+\delta_n)\ket{n} \end{equation} where $\mathcal{N}$ denotes a normalization constant and $\delta_n$ represents the perturbation introduced to $\ket{\psi_n}$ (noise term). Noise is immediately added to each amplitude of the state vector prior to normalization. The introduction of noise causes a deviation from the original value. Such Gaussian-distributed numerical perturbations are widely used in computational studies to probe the stability of quantum states, measurement outcomes, and phase-space representations under stochastic fluctuations, and to provide qualitative insight into noise sensitivity and robustness \cite{Weedbrook2012}.

White noise is commonly used in quantum information theory as an effective model to describe uniform random mixing arising from environmental imperfections. In this work, white noise is modeled using the depolarizing channel, which replaces the quantum state with a maximally mixed state with a certain probability. Mathematically,
\begin{equation}
\rho \longrightarrow \rho'=(1-p)\rho + p\frac{I}{d}
\end{equation} 
where \( p \) denotes a minor error (\( 0 \leq p \leq 1 \)), \( d \) represents the dimensions of the system space (\( 2^n \)), and \( \rho \) and \( \rho' \) indicate the original and noise-added density matrices, respectively. Equation (9) illustrates that with probability \(1-p\), the state remains unchanged, whereas with probability \(p\), the state is replaced by a maximally mixed state \(\frac{I}{d}\). White noise is not incorporated into each amplitude of the basis states; rather, the entire density matrix is combined with the maximally mixed state $\frac{I}{d}$. The depolarizing channel serves as a widely used benchmark noise model in quantum information theory for studying decoherence, loss of coherence, and robustness of quantum states under uniform environmental disturbances.

\subsection{Fidelity as a Measure of Noise Robustness}
The measure of fidelity plays an important role in quantum information theory. This is because fidelity quantifies the degree of similarity between the actual outcome and the desired state during an quantum operations. Fidelity usually depends on the initial state on which the operation or the noise is applied. Fidelity can also be interpreted as a similarity measure related to distance-like quantities in Hilbert space, with higher fidelity indicating greater closeness between quantum states. In general fidelity between a pair of density matrices $\rho$ and $\sigma$ is defined as\cite{Jozsa1994},
\begin{equation}
    F(\rho, \sigma) = (\mathrm{Tr} \sqrt{\sqrt{\rho}\sigma  \sqrt{\rho}})^2
\end{equation}
This is called Uhlmann-Jozsa fidelity, providing a unified definition for both pure and mixed quantum states. Table I summarizes the fidelity expressions based on the nature of the states being studied.

In this work, fidelity is employed as a quantitative metric to study the robustness of GHZ(3) and W(3) states under Gaussian-distributed perturbations and white noise, where it quantifies how close the noise state is to our ideal state \cite{Bennett1996,Carvalho2004}. The fidelity analysis complements the probability distributions and phase-space visualization based on Wigner functions. 

\begin{table}[h]
\centering
\caption{Fidelity expressions for the states under study}
\label{tab:fidelity_definitions}
\renewcommand{\arraystretch}{1.3}
\begin{tabular}{|c|p{3cm}|p{3.5cm}|}
\hline
\textbf{Case} & \textbf{Quantum states compared} & \textbf{Fidelity formula} \\
\hline
Pure--Pure &
$|\psi\rangle ,\, |\psi'\rangle$ &
$F = \left|\langle \psi \mid \psi' \rangle \right|^2$ \\
\hline
Pure--Mixed &
$|\psi\rangle ,\, \sigma$ &
$F = \langle \psi \mid \sigma \mid \psi \rangle$ \\
\hline
Mixed--Mixed &
$\rho ,\, \sigma$ &
\begin{tabular}[c]{@{}l@{}}
$F = \left( \operatorname{Tr}
\sqrt{ \sqrt{\rho}\, \sigma \, \sqrt{\rho} } \right)^2$
\end{tabular} \\
\hline
\end{tabular}
\end{table}

\subsection{Density Matrix Formalism and Purity}
When we say with complete definiteness that a system is in the quantum state $\ket{\psi}$, that kind of state is called a pure quantum state. There is no randomness in specifying the state of the system. Unlike pure states, mixed states cannot be represented by a single state vector in Hilbert space \cite{Holevo2012}.

The density matrix formalism allows calculations of probabilities and expectations without requiring knowledge of the exact state \cite{NielsenChuang,Breuer2002,Holevo2012}. The density operator is given as $\hat{\rho}$, defined mathematically as,
\[
\hat{\rho} = \sum_j p_j \ket{\psi}_j \bra{\psi}_j
\]

This density operator can be used to describe both pure and mixed states \cite{blum2012density}. For pure state one of the probabilities equals unity and the rest vanish,

\[
\hat{\rho} = \ket{\psi} \bra{\psi}
\]

For a quantum system in a orthonormal basis $\ket{i}$, the structure of the density matrix is

\begin{equation}
\hat{\rho} =
\begin{bmatrix}
\hat{\rho}_{11} & \hat{\rho}_{12} & \hat{\rho}_{13} & \cdots \\
\hat{\rho}_{21} & \hat{\rho}_{22} & \hat{\rho}_{23} & \cdots \\
\vdots          & \vdots          & \vdots          & \ddots
\end{bmatrix}.
\end{equation}

The diagonal elements $\hat{\rho}_{ii}$ represent the probability of finding the state in the eigenstate $\ket{i}$, while the off-diagonal elements $\hat{\rho}_{ij}$ contribute to the coherence \cite{Schlosshauer2004,Baumgratz2014}. They represent the phase relation between $\ket{i}$ and $\ket{j}$, resulting in a coherence term. Coherence allows quantum interference, entanglement, and computational advantage in quantum computing. So, loss of coherence means the system is losing the quantum properties and behaving classically.

 The purity of a quantum state is quantified by

\[
\gamma = \mathrm{Tr}(\rho^2),
\]
where $\gamma = 1$ corresponds to a pure state and $\gamma < 1$ indicates a mixed state.

The Gaussian-distributed amplitude perturbations implemented in this work act directly on the state amplitudes and produces a renormalized state vector. Consequently, the resulting quantum state can still be expressed as a single ket $|\psi'\rangle$, and the corresponding density matrix $\rho' = |\psi'\rangle\langle\psi'|$ satisfies $\mathrm{Tr}(\rho'^2)=1$. The noisy state therefore remains pure, and the observed degradation of multipartite correlations reflects an effective decoherence rather than a genuine mixed-state evolution arising from system–environment interactions . In contrast, white noise introduces a statistical mixture with the maximally mixed state, leading to $\mathrm{Tr}(\rho^2) < 1$, and represents true decoherence accompanied by loss of purity.

\subsection{Wigner Function Representation of Quantum States}
The Wigner function is a mathematical tool that represents quantum states in phase space (position and momentum). For a quantum state described by the density operator $\rho$, the Wigner function, contains all the information needed to reconstruct the quantum state \cite{Hillery1984}. For continuous-variable systems, the Wigner function is defined as

\begin{equation}
W(x,p) = \frac{1}{\pi\hbar}\int_{-\infty} ^{\infty} dy \quad e^{\frac{2ipy}{\hbar}} \bra{x-y}\rho \ket{x+y}
\end{equation}

This is analogous to how probability distributions describe classical systems. Unlike those true probabilities, the Wigner function can take negative values \cite{Kenfack2004,Ballicchia2019Investigating}.While it remains normalized,

\begin{equation}
\int dx\,dp \, W(x,p) = 1
\end{equation}

The negative reflects the quantum features such as superposition and entanglement \cite{Kenfack2004,Ballicchia2019Investigating}. This is called quasiprobability distribution because it is not always positive \cite{Hillery1984}. Negative regions in the Wigner functions indicate non-classicality and are used as a signature of quantum behavior \cite{Kenfack2004,Ballicchia2019Investigating}. Integrating Wigner functions over momentum (or position) gives the correct quantum probability distribution for position (or momentum). 

\begin{equation}
\begin{aligned}
\int _{-\infty} ^{\infty} dp W(x,p) = \bra{x} \rho \ket{x}, \\
\int _{-\infty} ^{\infty} dx W(x,p) = \bra{p} \rho \ket{p}
\end{aligned}
\end{equation}

For finite-dimensional systems like multi-qubit quantum systems, the Wigner function is defined using the phase-point operator $\hat{A}(\alpha)$ \cite{Gross2006,Tilma2016,millen2023generalized,Leiner2017Wigner},

\begin{equation}
W(\alpha) = \mathrm{Tr[\hat{\rho} \hat{A}(\alpha)]}
\end{equation}

where $\alpha$ points to the labels in discrete phase space. For multipartite systems, the Wigner function captures global quantum correlations \cite{Vesperini2023Correlations}. Highly entangled states, such as GHZ and W states, exhibit interference patterns and significant negativity in their Wigner representations \cite{Rahman2021Probing}.

In this study, the Wigner function framework is used to investigate the behavior of GHZ and W state under white and Gaussian-distributed amplitude perturbations. The evolution of phase-space structures and the degradation of negativity provide direct insight into the robustness of multipartite entanglement in noisy quantum environments \cite{Hawary2024Navigating}.

\subsection{Spin Wigner Representation with Equal-Angle Projection}
To visualize tripartite GHZ and W states under Gaussian and white noise, we use the spin Wigner function, and equal angle projection is used to reduce dimensions.

The spin Wigner function is the phase space visualization for finite-dimensional quantum systems. Here, phase space is the surface of the Bloch sphere parametrized by $(\theta, \phi)$, instead of position and momentum.

Instead of measuring each qubit in different bases, the same measurement direction is applied to every qubit in the system. This kind of projection is called equal angle projection and is useful for entangled states, as it preserves and highlights global quantum correlations.

For N qubits, the full spin Wigner function is given by

\[
W(\theta_1,\phi_1 ; \theta_2,\phi_2 ; ...... ;\theta_n,\phi_n)
\]

i.e., 2N continuous variables, which is very difficult to visualize directly \cite{Leiner2017Wigner}.

Hence, we set all the qubits to the same point in the Bloch sphere resulting in equal angles,

\[
\begin{aligned}
\theta_1 = \theta_2  = ....... = \theta_n \\
\phi_1 = \phi_2 = .... = \phi_n
\end{aligned}
\]
and the Wigner function becomes,
\begin{equation}
W_{EA} = \mathrm{Tr[\rho \pi(\theta, \phi)^{\otimes N}]}
\end{equation}

where $(\theta, \phi)$ is the direction on the Bloch sphere, and $\pi(\theta, \phi)$ is the mathematical probe that checks whether a qubit is aligned or anti-aligned with the direction $(\theta, \phi)$. It is a parity-like operator \cite{Leiner2017Wigner}.

\begin{equation}
\pi(\theta, \phi) = \frac{1}{2} (\mathbb{I} + \sqrt{3} \mathbf{n}(\theta, \phi).\boldsymbol{\sigma})
\end{equation}

where $\frac{1}{2}$ is the normalization constant, $\boldsymbol{\sigma}$ is the Pauli operator, and the factor $\frac{\sqrt{3}}{ 2}$ is derived from Stratonovich-Weyl conditions, namely, normalization, rotation covariance, self-duality, and completeness over the sphere. Eigenvalue of $\pi(\theta, \phi)$ is $\frac{1}{2}(\mathbb{I} \pm \sqrt{3}) $, which means that one eigenvalue is always negative and is the source of Wigner negativity.

The spin Wigner function at each point represents a real value that encodes information about the quantum state under that specific direction. Positive regions correspond to classical-like behavior, while the negative regions specify pure quantum interference. Zero crossings indicate a balance between classical and quantum correlations. The presence of negativity is therefore a direct signature of quantum coherence and entanglement \cite{Ballicchia2019Investigating,Rahman2021Probing}.

\section{Comparative Analysis of GHZ(3) and W(3) States under Gaussian and White Noise}
Figure 1 shows the probability distribution of the ideal three-qubit GHZ(3) state and W(3) state. As previously described in equations (3) and (4), the GHZ(3) state exhibits equal probability for the computational basis states $\ket{000}$ and $\ket{111}$, with the probabilities of other computational bases being zero. While in the case of the W(3) state, the probabilities are equally distributed among $\ket{001}$, $\ket{010}$, and $\ket{100}$, reflecting the single-excitation symmetry of the state.
\begin{figure*}
\xincludegraphics[width=0.35\linewidth, label={a)}]{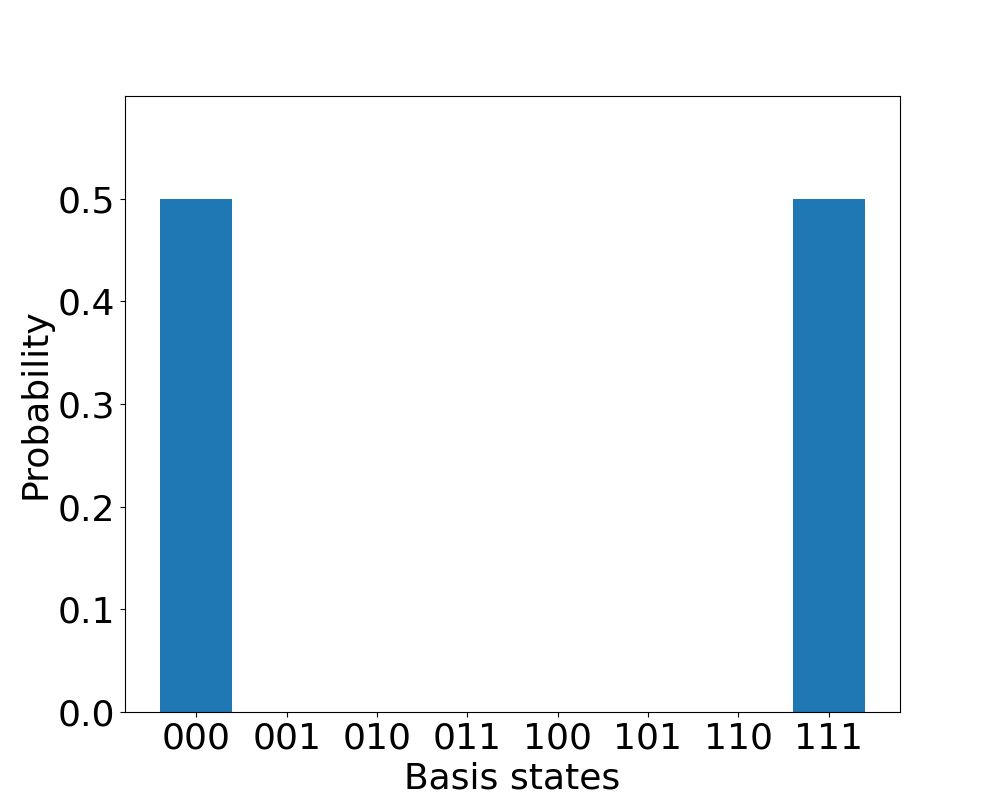}
\xincludegraphics[width=0.35\linewidth, label={b)}]{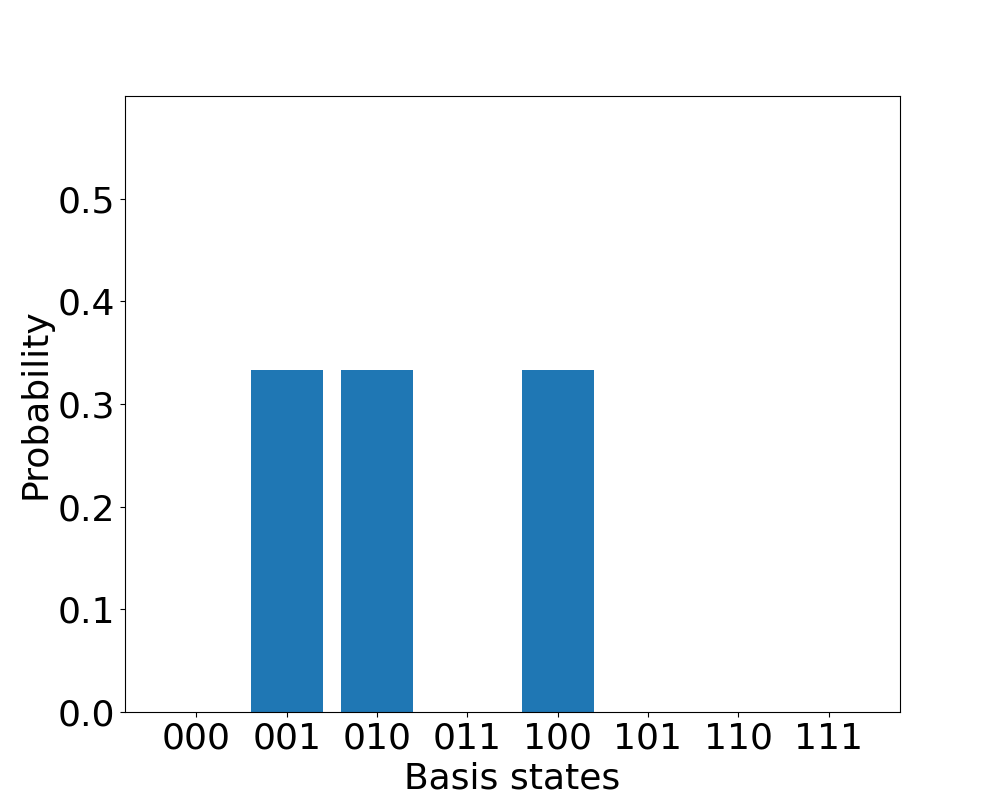}

    \caption{  The probability distribution of the (a) ideal GHZ(3) state across the basis states exhibiting perfect correlations between $\ket{000}$ and $\ket{111}$. (b) the ideal W(3) state across the basis states $\ket{001}$, $\ket{010}$, and $\ket{100}$ with equal probabilities.}

\end{figure*}

\subsection{Probability Distribution under Noise}

Figure 2 illustrates the results of GHZ(3) and W(3) states under gaussian-distributed amplitude perturbation with varied  strengths $\sigma=0.1$ and $\sigma=1.0$. The Gaussian-distributed amplitude perturbation was introduced using the $\texttt{add$\textunderscore$random$\textunderscore$noise(psi, m=0.0, st)}$, with $st \in [0.1,1.0]$ \cite{tuan2021tqix}. As standard deviation increases, the strength of noise also increases.

At the low noise range $\sigma=0.1$, the probability began to spread among the other basis states, but our initial basis states kept dominating. This explains that at low noise levels, the probability level is highly concentrated on the primary basis states, indicating the characteristic of ideal GHZ(3) and W(3) states. 

As the noise level increased and reached (a) and (b) $\sigma=0.4$, it is seen that the probability is redistributed among other computational basis states apart from our initially dominant basis states $\ket{000}$, $\ket{111}$,  $\ket{001}$, $\ket{010}$, and $\ket{100}$ for GHZ(3) and W(3) states, respectively \cite{Rahman2021Demonstration}. As we reach (c) and (d) $\sigma=1.0$, we note that the probability is redistributed completely, and this reflects the suppression of ideal states. This redistribution of probability indicates the loss in coherence and multipartite entanglement \cite{Zhou2019Entanglement,Laskowski2014Highly}.

\begin{figure*}
\centering
\begin{tabular}{ccccccc}
\xincludegraphics[width=0.35\linewidth, label={a)}]{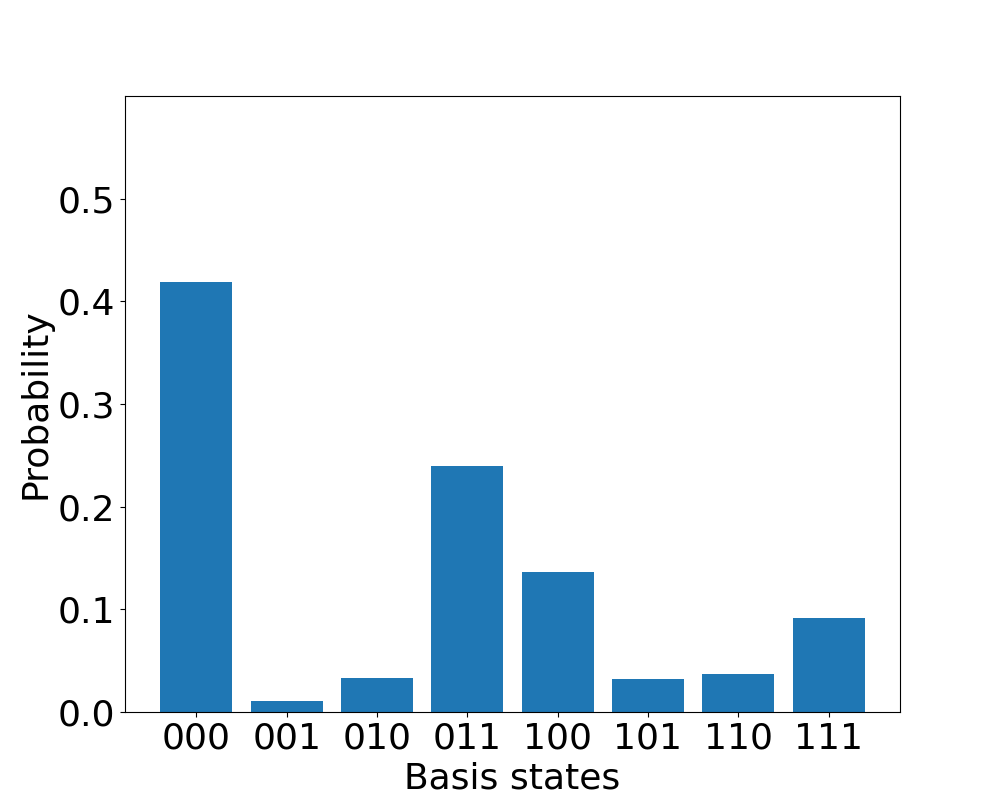} &
\xincludegraphics[width=0.35\linewidth, label={b)}]{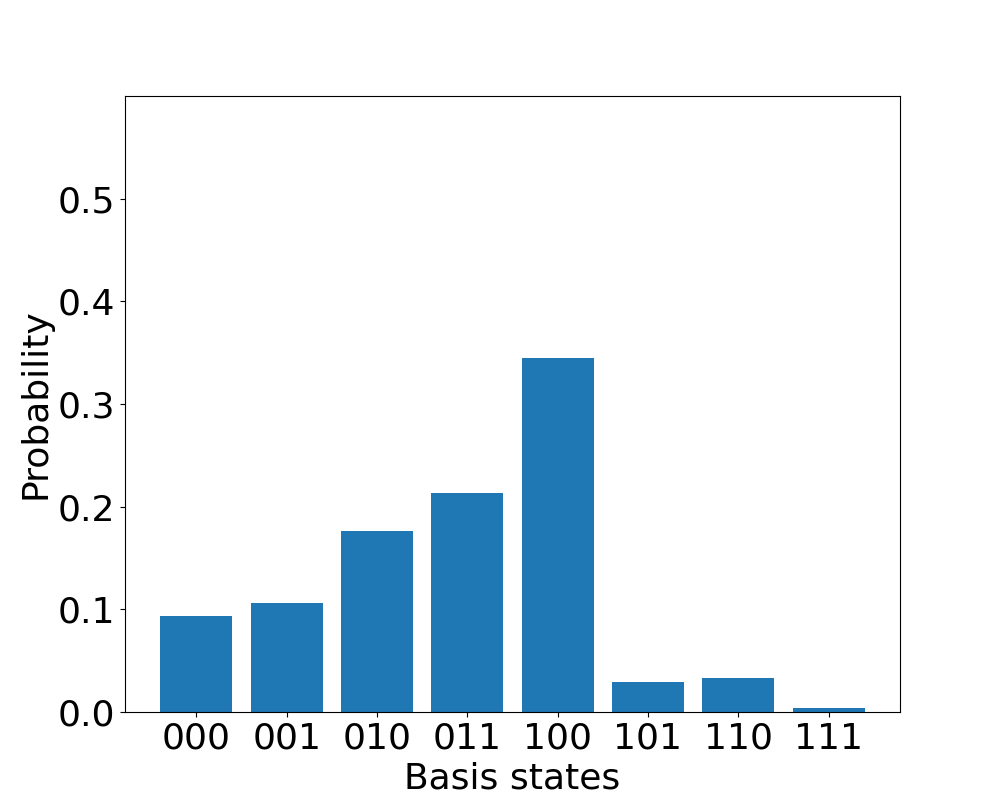} \\
\xincludegraphics[width=0.35\linewidth, label={c)}]{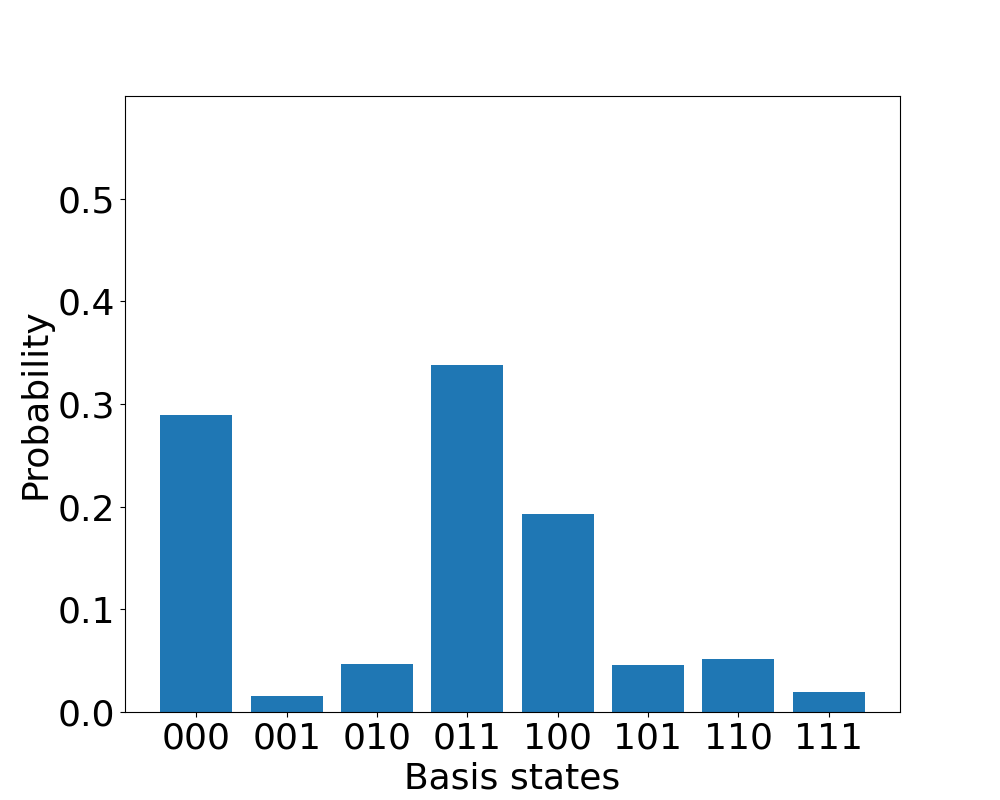} &
\xincludegraphics[width=0.35\linewidth, label={d)}]{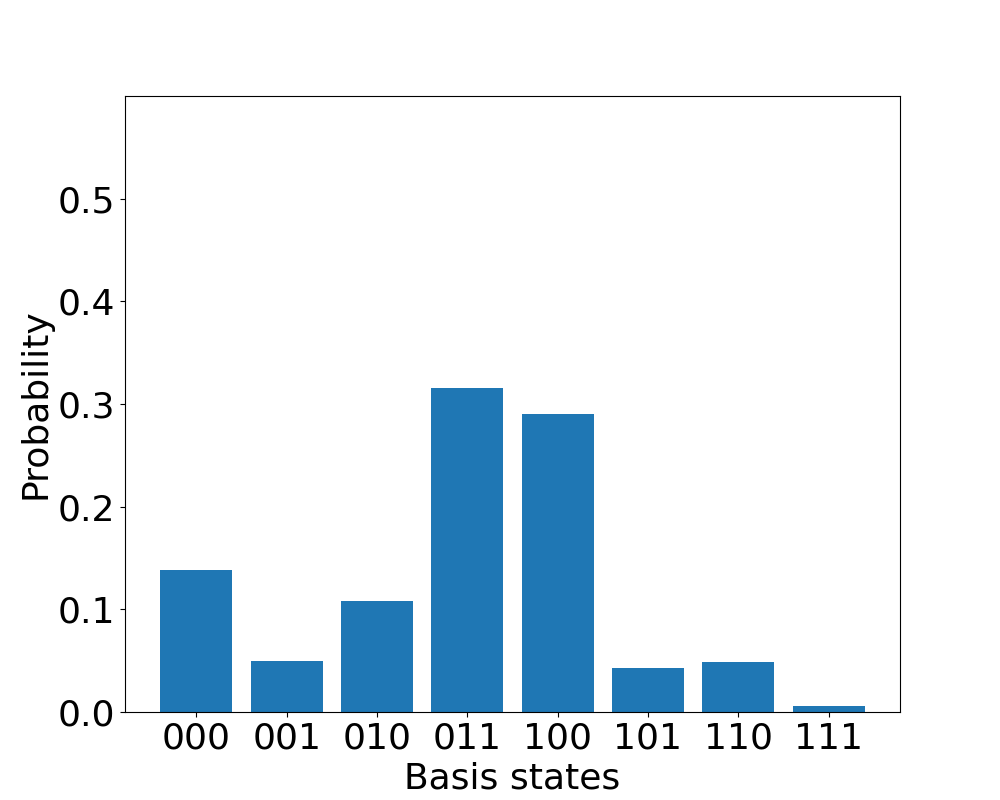}
\end{tabular}
    \caption{Comparative analysis of GHZ(3) and W(3) states under gaussian-distributed amplitude perturbation for increasing noise strength $\sigma$: panels (a) and (b) show the probability distribution of GHZ(3) and W(3) states, respectively, at noise level $\sigma=0.4$, showing a noticeable probability redistribution into other basis states; panels (c) and (d) show the strong noise level $\sigma=1.0$ of GHZ(3) and W(3) states, respectively, showing suppression of the ideal states.}
\end{figure*}

Though both states result in degradation of the coherent superposition, a clear contrast is seen in their robustness.  The GHZ(3) state shows a rapid degradation in their initial computational basis as the noise value increases, while it was seen that for the W(3) state, it retains a higher probability for the initial basis states even at stronger noise levels till $\sigma=0.8$ \cite{Rahman2021Quantum}.

Figure 3 illustrates the results of GHZ(3) and W(3) states under white noise with varied noise strength $p=0.4$ and $p=1.0$. The white noise is introduced using \cite{tuan2021tqix} $\texttt{add$\textunderscore$white$\textunderscore$noise(rho, p)}$, with $p \in [0.1,1]$. As p is increased, the strength also increases simultaneously. 

Similar to gaussian-distributed amplitude perturbation, at a low noise range $p=0.1$, the probability is spread among the other basis states uniformly, but our primary basis states kept dominating. This explains that at low noise levels, the probability level is highly concentrated on the primary basis states, indicating the characteristic of ideal GHZ(3) and W(3) states. 

As the noise level increases to (a) and (b) $p=0.4$, it is seen that the probability is redistributed uniformly among other computational basis states apart from our initially dominant basis states $\ket{000}$, $\ket{111}$, $\ket{001}$, $\ket{010}$, and $\ket{100}$ for GHZ(3) and W(3) states, respectively. As the noise strength reaches its maximum at (c) and (d) $p=1.0$, we note that the probability is redistributed completely uniformly, and this reflects the suppression of ideal states. This redistribution of probability indicates the loss in coherence and entanglement \cite{Zhou2019Entanglement,Laskowski2014Highly,Mor-Ruiz2023Influence}.
\begin{figure*}
\xincludegraphics[width=0.35\linewidth, label={a)}]{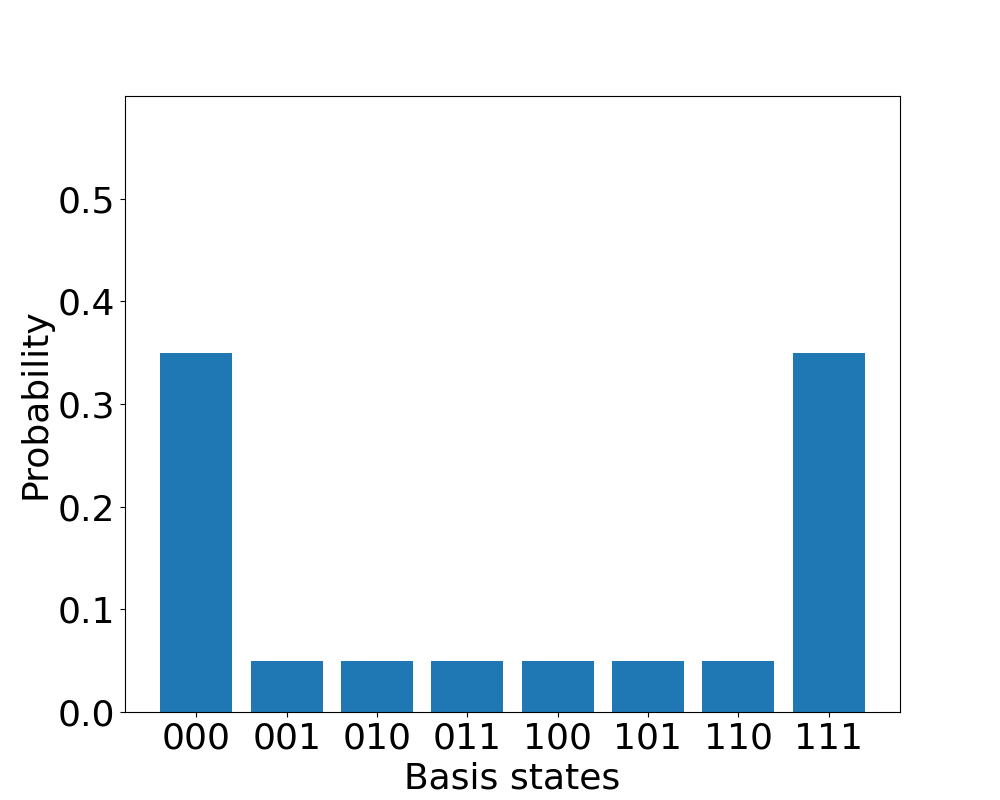} 
\xincludegraphics[width=0.35\linewidth, label={b)}]{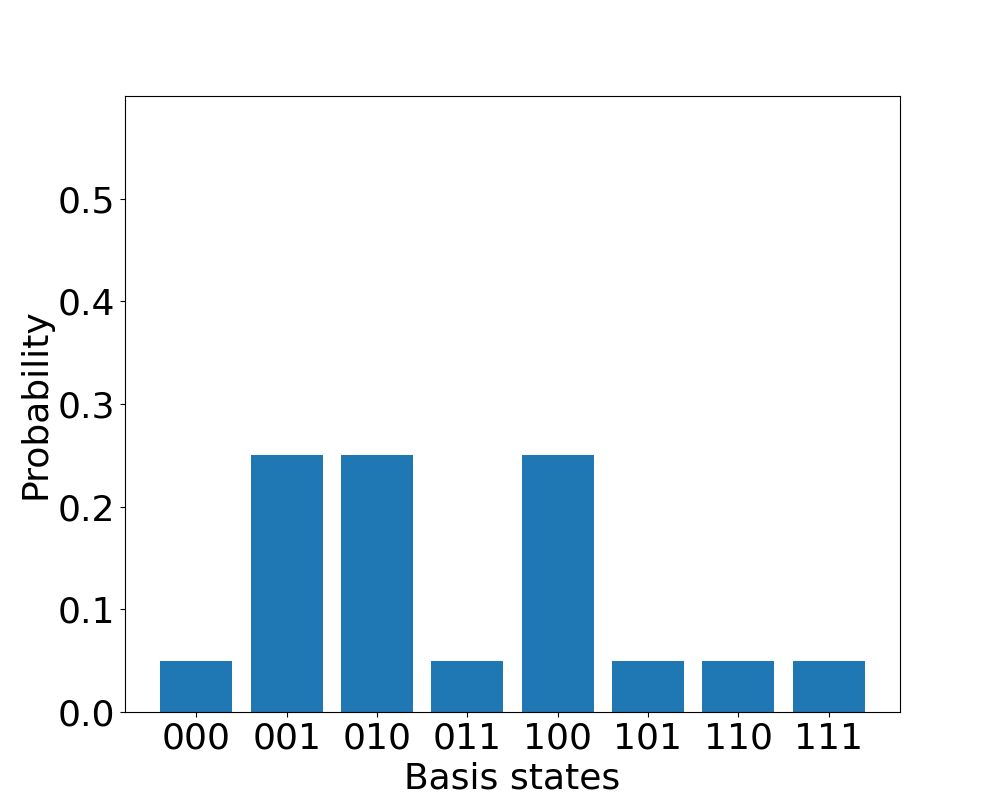} 
\xincludegraphics[width=0.35\linewidth, label={c)}]{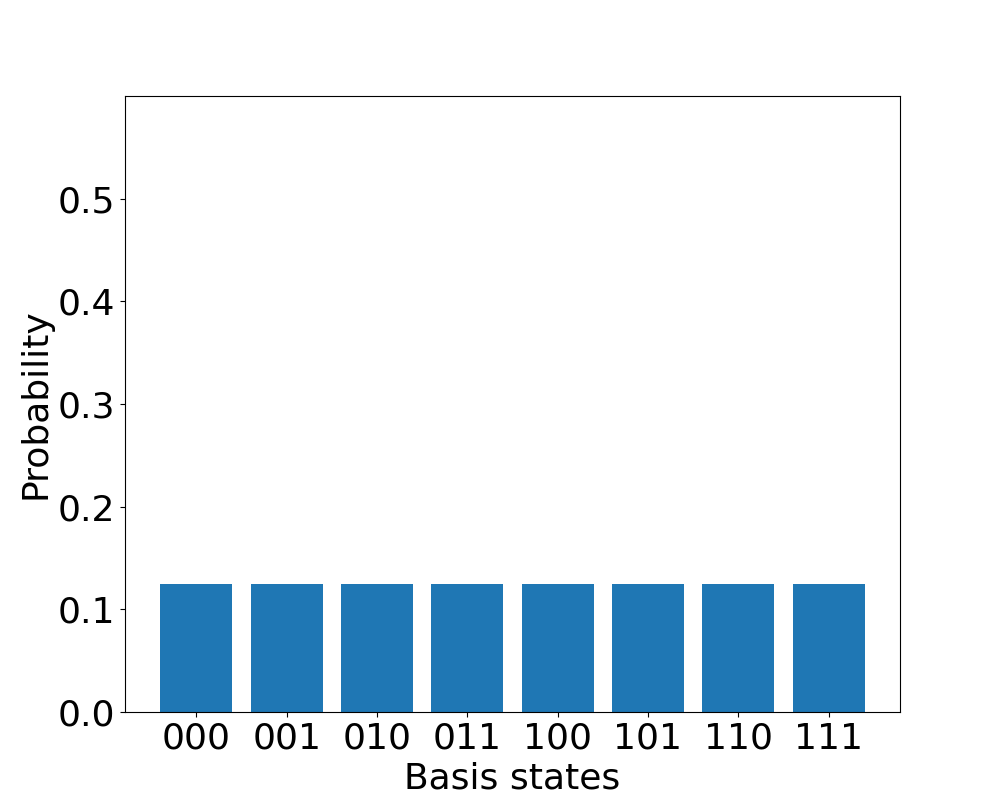} 
\xincludegraphics[width=0.35\linewidth, label={d)}]{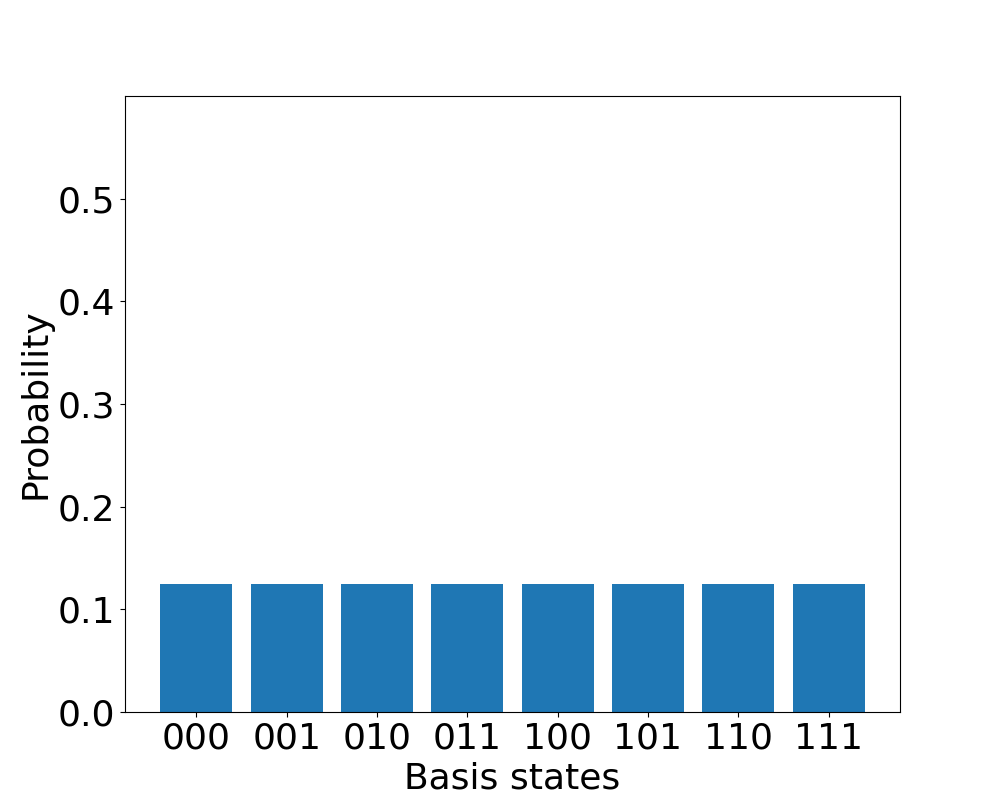}
    \caption{Comparative analysis of GHZ(3) and W(3) states under white noise for increasing noise strength $p$: panels (a) and (b) show the probability redistribution of GHZ(3) and W(3) states, respectively, at noise level $p=0.4$, showing a noticeable uniform probability redistribution into other basis states; panels (c) and (d) show the strong noise level $p=1.0$ of GHZ(3) and W(3) states, respectively, showing suppression of the ideal states.} 
\end{figure*}

 Though both GHZ(3) and W(3) states resulted in uniform redistribution of probabilities under white noise, it is seen that the GHZ(3) state showed a gradual redistribution of the probabilities compared to W(3) states \cite{Rahman2021Quantum,Rahman2021Dynamics,Kenfack2018Decoherence}. This can be explicitly seen in panels (a) and (b).

\subsection{Fidelity Analysis}

Figure 4 illustrates the comparative behavior of the Uhlmann–Jozsa fidelity for three-qubit GHZ(3) and W(3) states under Gaussian and white noise. Panels (a) and (b) present the effect of gaussian-distributed amplitude perturbation and white noise on GHZ(3) and W(3) states, respectively \cite{Espoukeh2014Quantum,Dai2018Experimentally,Ghoshal2023All,Mandarino_2016}. It is observed that, under both Gaussian and white noise, the fidelity curves of the two tripartite entangled states overlap almost over the entire range of noise strength. This overlap indicates a nearly identical decay of fidelity for GHZ(3) and W(3) states when subjected to the same noise model \cite{Mandarino_2016,Dai2018Experimentally}. Consequently, this behavior suggests that fidelity does not distinguish how different multipartite entanglement structures respond under similar noise conditions \cite{Ghoshal2023All}. On the other hand, panels (c) and (d) compare GHZ(3) and W(3) states under Gaussian and white noise, respectively. It is noted that under gaussian-distributed amplitude perturbation, both the tripartite entangled states decay faster at the initial low noise levels, and as the noise level increases, the fidelity decay slows down and eventually reaches a finite non-zero value, while under white noise, both the states decay almost linearly at the same rate, indicating the uniform degradation.

Hence, from the nearly overlapping fidelity curve of the states under the same noise models, we conclude that, though fidelity provides a quantitative measure of the overall degradation of quantum states under noise conditions, it doesn't serve as a qualitative metric to visualize the states and doesn't certify quantum properties. The qualitative differences in the phase-space structures and the non-classical features of the different entanglement types are examined and visualized in detail using the equal-angle spin Wigner function in the upcoming sections.
\begin{figure*}
\centering
\begin{tabular}{ccccccc}
\xincludegraphics[width=0.35\linewidth, label={a)}]{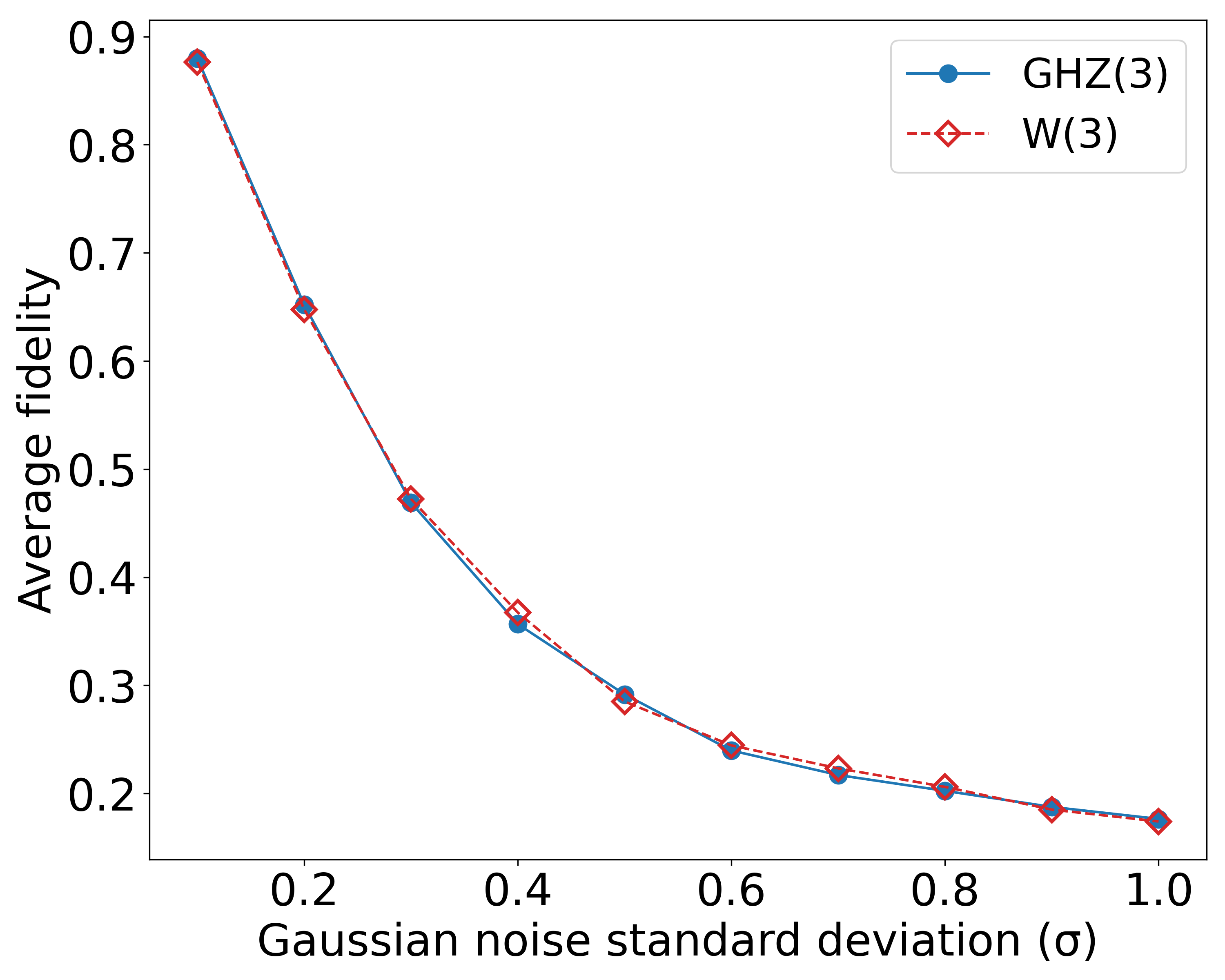} &
\xincludegraphics[width=0.35\linewidth, label={b)}]{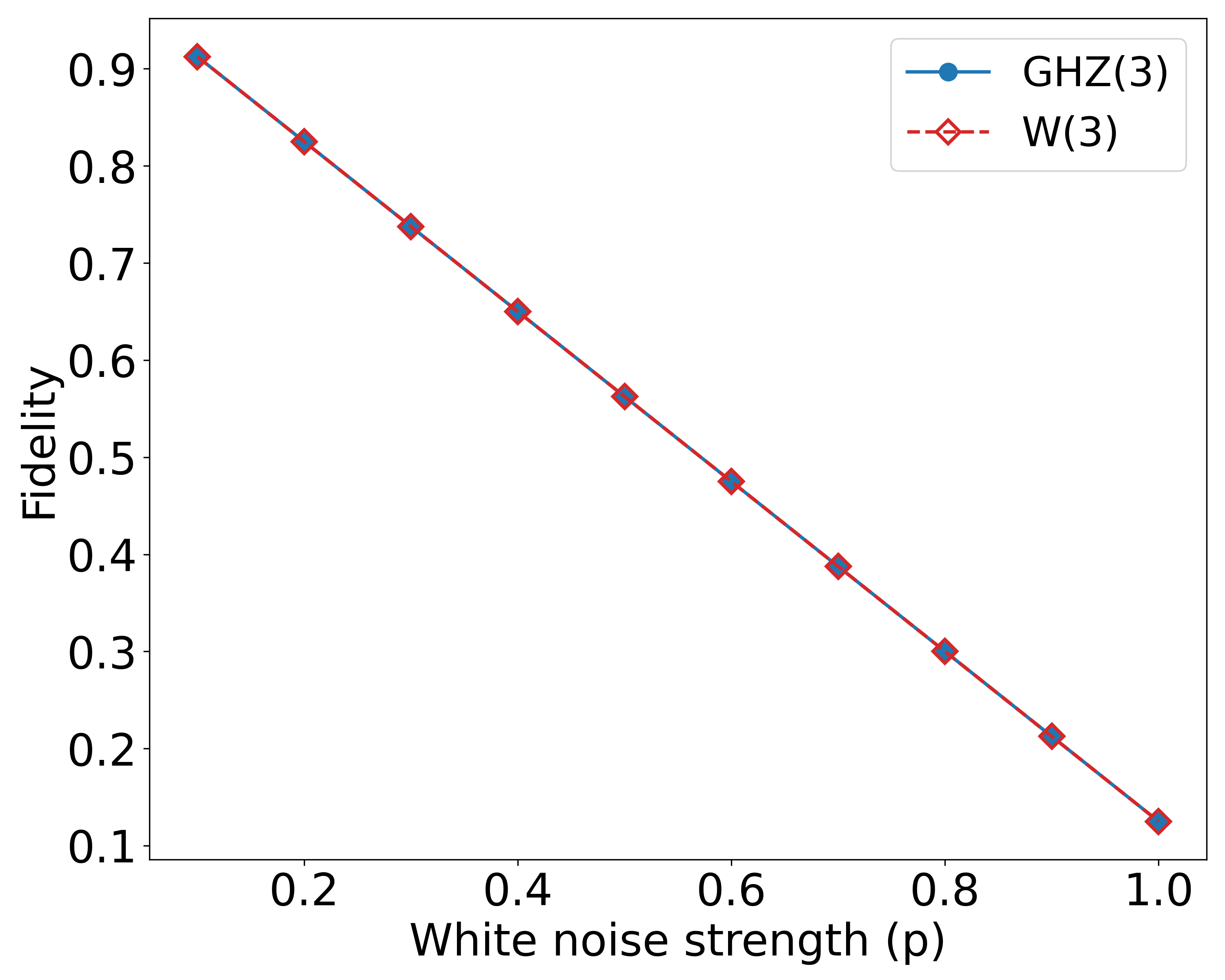} \\
\xincludegraphics[width=0.35\linewidth, label={c)}]{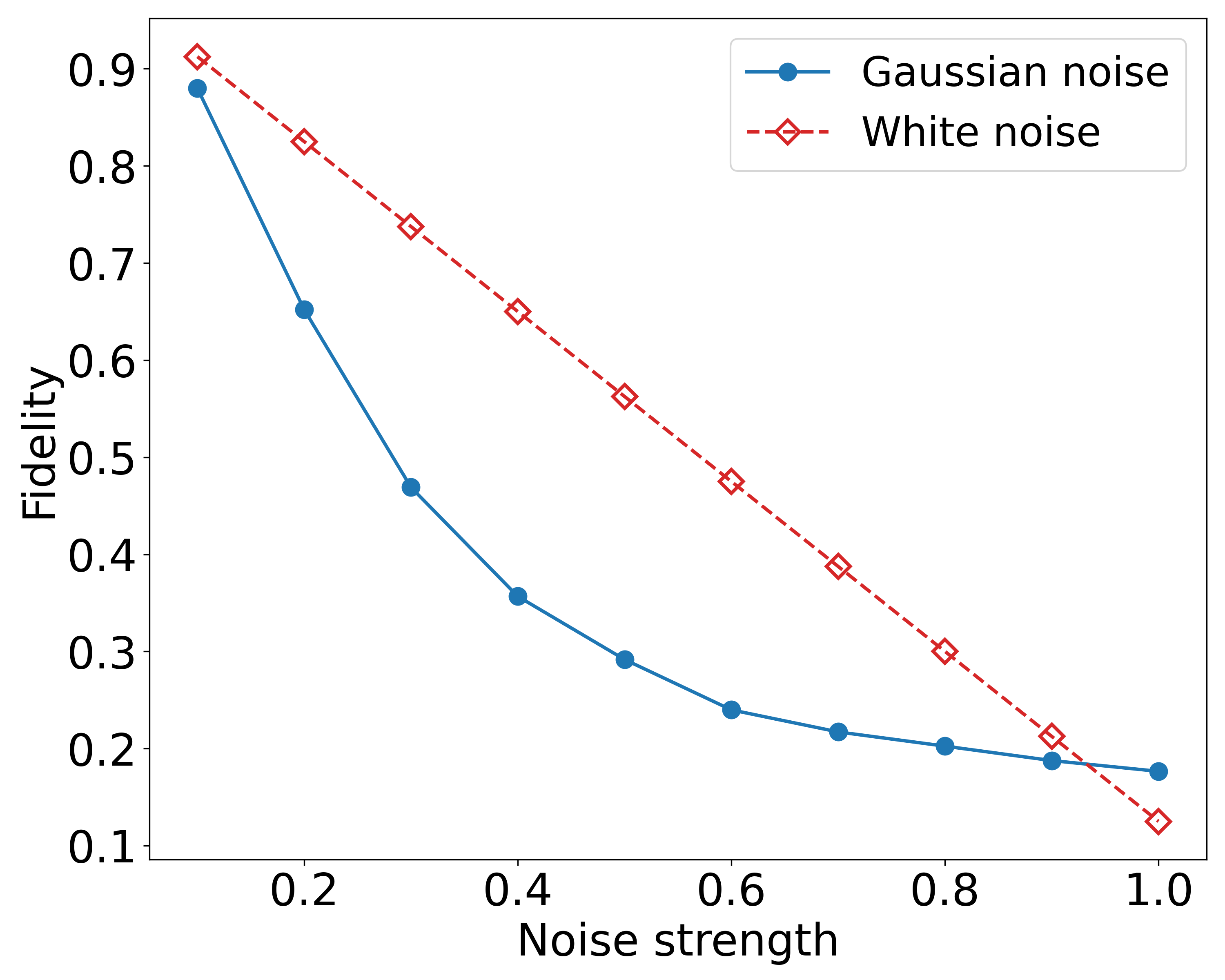} &
\xincludegraphics[width=0.35\linewidth, label={d)}]{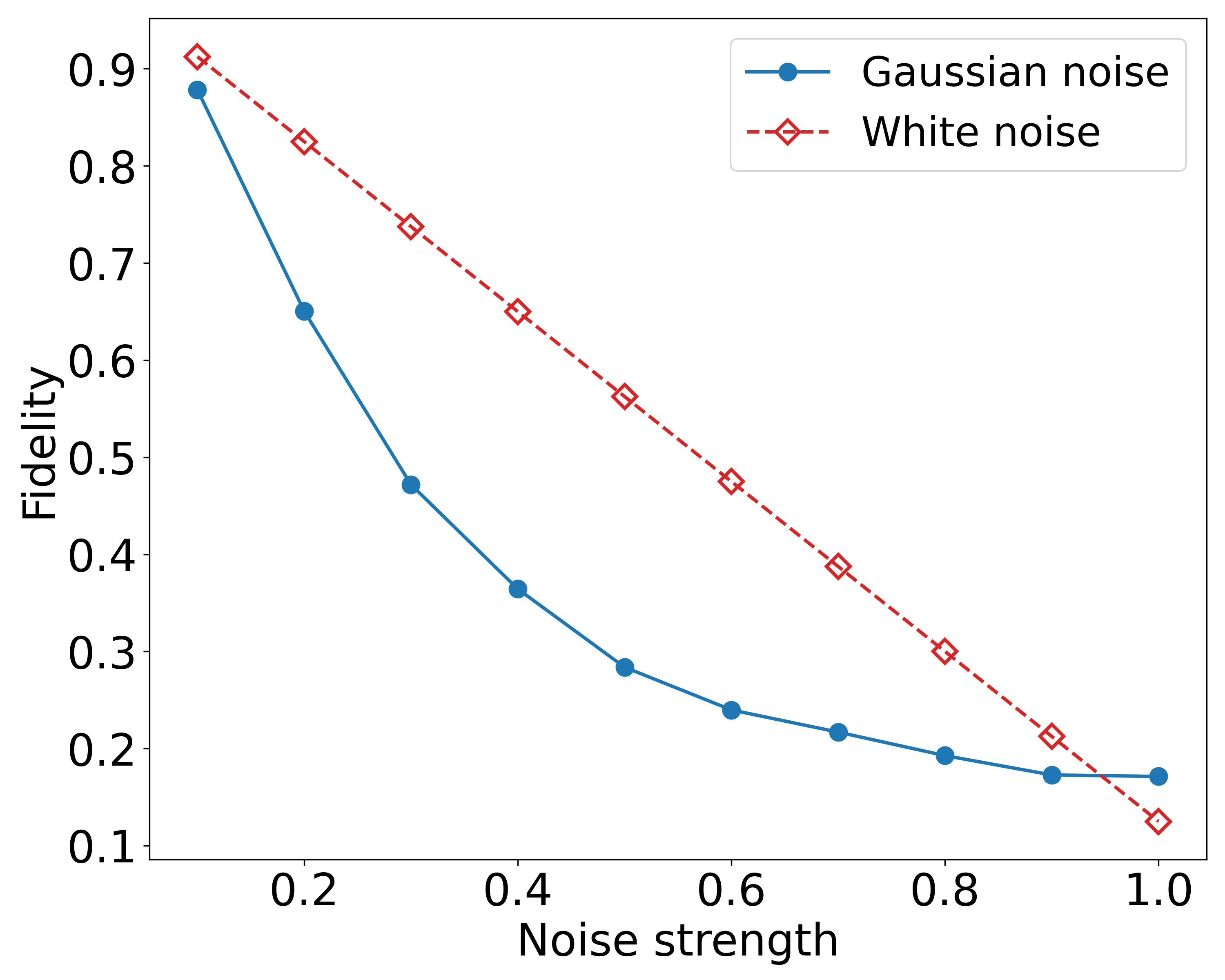}

\end{tabular}
    \caption{ Fidelity analysis of three-qubit GHZ and W states under Gaussian and white noise. Panel (a) and (b) show the comparison of fidelity of GHZ(3) and W(3) under gaussian-distributed amplitude perturbation and white noise, respectively; panel (c) compares the fidelity of the GHZ(3) state under gaussian-distributed amplitude perturbation; and panel (d) compares the fidelity of the W(3) state under Gaussian and white noise, with red and blue plots corresponding to GHZ(3) and W(3), respectively. }
\end{figure*}

\section{Wigner Function–Based Characterization of GHZ(3) and W(3) States}
In Figure 5, the panels (a) and (b) correspond to the tripartite GHZ state. The two-dimensional contour plot in panel (a) represents interference patterns along the azimuthal angle, represented by alternating red and blue regions \cite{Ciampini2017Wigner,Sánchez-Soto2025Phase}. The red regions correspond to the positive values of the Wigner function, specifying classical-like behavior, while the blue regions contribute to negative values corresponding to quantum characteristics \cite{Ziane2019The,Walschaers2021Non-Gaussian}. The presence of blue lobes corresponds to strong coherence between $\ket{000}$ and $\ket{111}$ components of the GHZ state. This is further visualized in the three-dimensional surface plot in panel (b). The sharp valleys and the peak specify negative and positive contributions to the Wigner function, respectively.
\begin{figure*}
\centering
\begin{tabular}{ccccccc}
\xincludegraphics[width=0.3\linewidth, label={a)}]{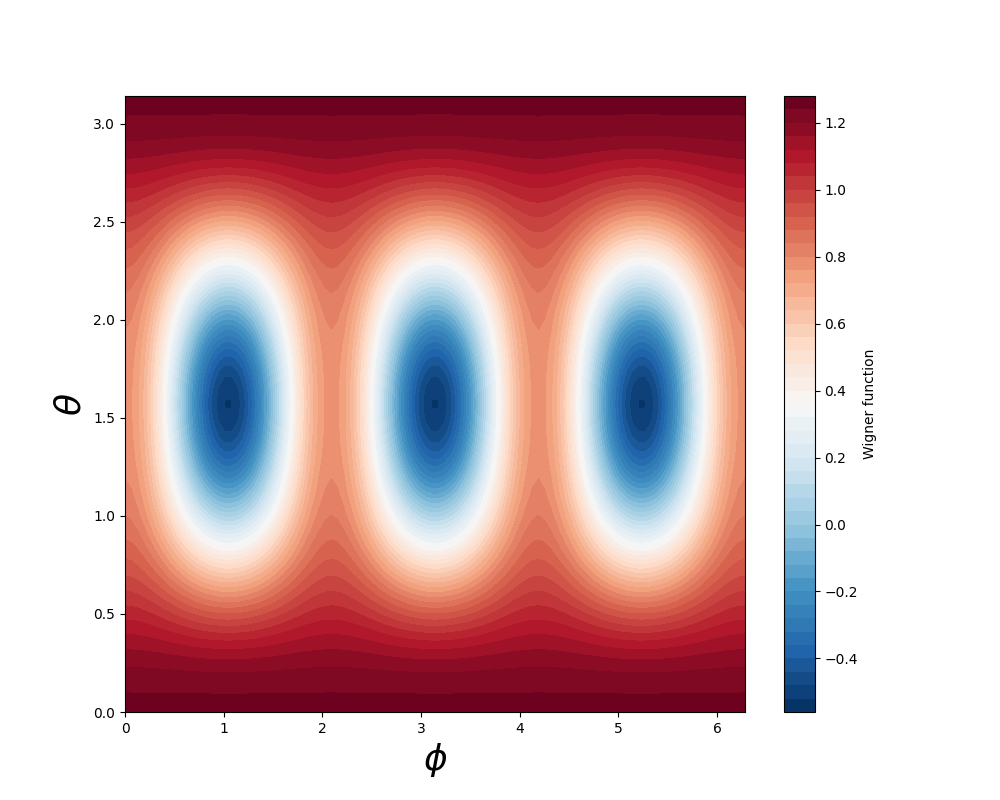} &
\xincludegraphics[width=0.35\linewidth, label={b)}]{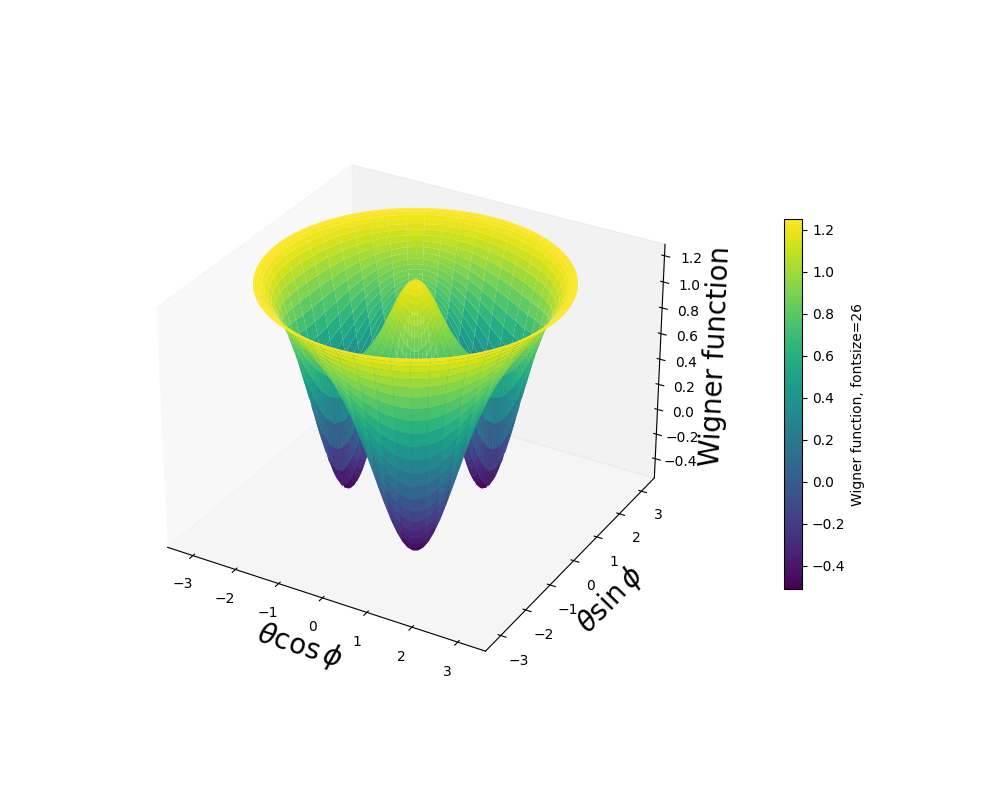} \\
\xincludegraphics[width=0.3\linewidth, label={c)}]{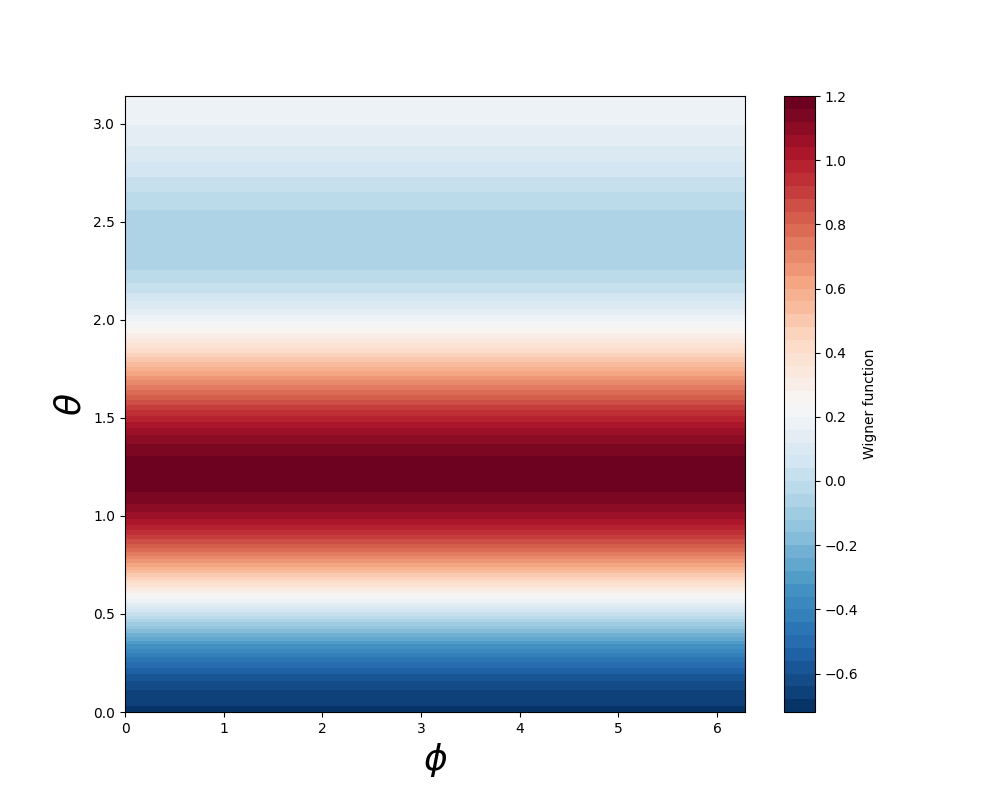} &
\xincludegraphics[width=0.35\linewidth, label={d)}]{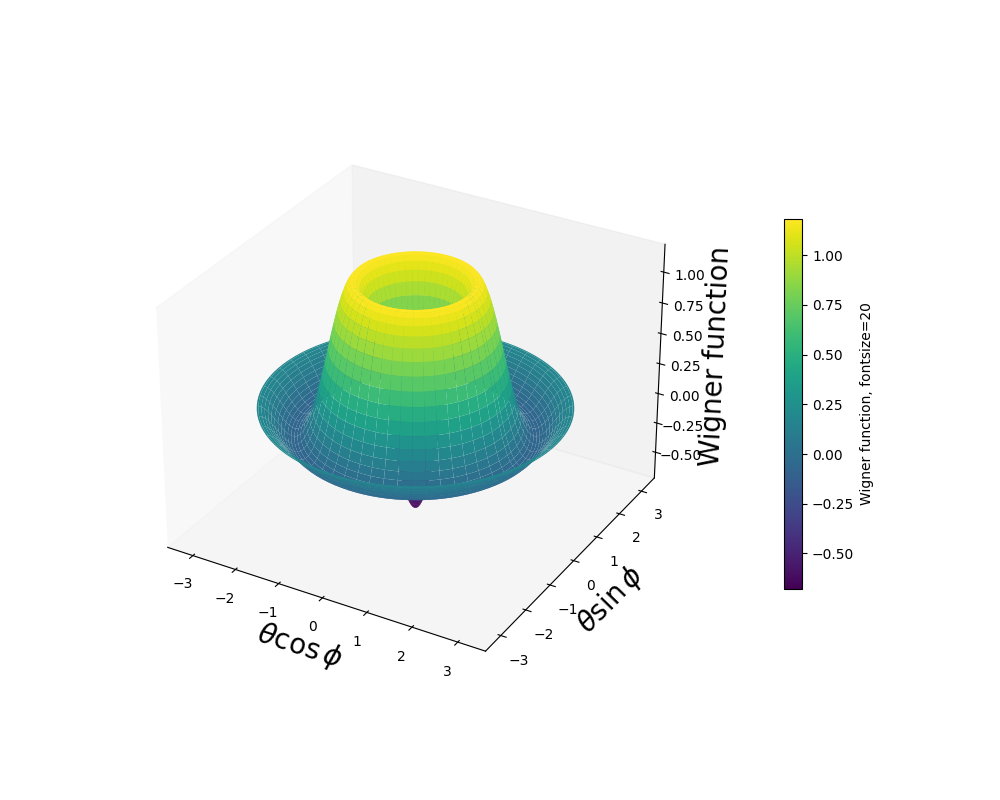}
\end{tabular}
    \caption{ Equal-angle Spin Wigner Function for three-qubit GHZ and W states. Panels (a) and (c) represent two dimensional contour plots of equal angle spin Wigner function for GHZ and W states respectively plotted over Bloch sphere with $\theta \in [0,\pi]$ and $\phi \in [0,2\pi]$. Panels (b) and (d) specify the corresponding three-dimensional surface plots obtained by mapping the angular coordinates $(\theta, \phi)$ to Cartesian coordinates $(\theta \cos\phi, \theta \sin\phi)$, with the Wigner function value along the vertical axis. The color scale indicates the magnitude and sign of the Wigner function.}
\end{figure*}
Panels (c) and (d) illustrate the tripartite W state. Panel (c) depicts the two-dimensional contour plot of the W(3) state, which, in contrast to the GHZ(3) state, shows a smooth band-like structure corresponding to local entanglement/lack of global phase correlations compared to the GHZ(3) state. Similar to the case of the GHZ(3) state, red and blue regions correspond to positive and negative Wigner values, respectively. The weaker blue regions near the poles correspond to limited negativity and prove the lower degree of phase-dependent interference compared to the GHZ(3) state. Panel (d) corresponds to further visualization of the same in a three-dimensional surface plot, where the ring-like region corresponds to the single excitation level of the W(3) state, while the small blue region corresponds to local entanglement. 

\subsection{Noise-Induced Modifications of the Wigner Function}
In the case of gaussian-distributed amplitude perturbation, the perturbations are added directly to the amplitudes of the quantum state before normalization randomly \cite{Zhang2022From}. On the other hand, the white noise is implemented on the density matrix of the pure state, which is then mixed with the maximally mixed state \cite{Abd-Rabbou2019Wigner}. Hence, the resulting density matrix is an ensemble-averaged state, and the respective Wigner function is automatically averaged.

Figure 6 illustrates the equal-angle spin Wigner function of the three-qubit GHZ state under Gaussian and white noise \cite{Abd-Rabbou2019Wigner,Zhang2022From}. Panel (a) corresponds to GHZ(3) under gaussian-distributed amplitude perturbation, where the noise is applied directly to the amplitudes of the quantum state randomly corresponding to a single realization. (b) specifies GHZ(3) under white noise. In lower noise levels, under both Gaussian and white noise, the Wigner function shows pronounced interference structure and negative regions, which are the clear signature of non-classical correlations and genuine multipartite entanglement \cite{Walschaers2021Non-Gaussian,Ziane2019The}. As noise strength increases, these quantum features begin to degrade. Under gaussian-distributed amplitude perturbation (a), the state exhibits gradual deformation in the phase-space structure, indicating the degradation of quantum features like entanglement, interference, and correlations \cite{Zhang2022From}. Since the plots correspond to a single noisy realization, they capture the immediate impact of noise on the phase-space representation of the state. The ensemble-averaged Wigner function for both GHZ(3) and W(3) states under gaussian-distributed amplitude perturbation is shown in figure 8. In contrast, white noise leads to more uniform suppression of the structure, rapidly washing out both positive and negative regions of the Wigner function and driving the state toward a whole classical-like phase-space distribution \cite{Abd-Rabbou2019Wigner}. At the maximal white noise limit ($p=1.0$), the Wigner function of the W(3) state becomes completely flat. This behavior is consistent with the theoretical description presented in the previous section, where white noise drives the system toward a maximally mixed state \cite{Abd-Rabbou2019Wigner,Zhang2022From}. In this regime, all quantum coherence and interference features disappear, resulting in uniform distribution that is independent of $(\theta, \phi)$.

Figure 7 presents the equal-angle spin Wigner function of the three-particle W state (a) under Gaussian and (b) white noise. While both GHZ(3) (figure 6) and W(3) states exhibit quantum phase-space features at low noise levels, their response is totally different. This is due to their distinct entanglement structures \cite{De2003Multiqubit,Ciampini2017Wigner}. For Gaussian-distributed amplitude perturbations, similar to the GHZ(3) state, the noise is applied directly to the state amplitude. However, in contrast to GHZ(3), the Wigner function of the W(3) state shows a smoother and delocalized band-like structure\cite{Lalita2023Harnessing}. This reflects the distributed type entanglement of the W(3) state \cite{De2003Multiqubit}. As the noise strength increases, the phase space deforms more gradually than the GHZ(3) state. This indicates the comparatively higher robustness of the W(3) state \cite{De2003Multiqubit}. Under white noise, at intermediate noise levels, the decay is much slower than the GHZ(3) state. Similar to the GHZ(3) state at the maximal limit ($p=1.0$), the Wigner function is completely constant \cite{De2003Multiqubit}.

Figure 8 presents the ensemble-averaged spin Wigner function. This captures the response of the averaged phase-space structure of GHZ(3) and W(3) states under the effect of amplitude perturbations \cite{Zhang2022From}. This ensemble-averaged structure removes the random fluctuations formed in the previous cases [fig (6) and (7)] \cite{Zhang2022From}. At low noise strength, the phase space structure shows the non-classical features of GHZ(3) and W(3) states \cite{Walschaers2021Non-Gaussian}. As the noise strength increases, the interference patterns and the negative regions begin to suppress, leading to a smoother and classical-like structure \cite{Ziane2019The}. As the noise strength reaches the maximum level, both GHZ(3) and W(3) show a constant phase-space structure. This indicates that, in the case of ensemble-averaged gaussian-distributed amplitude perturbation, the state becomes mixed, unlike in the case of the realization-dependent case \cite{Zhang2022From}. For the case of the W(3) state, under increasing gaussian-distributed amplitude perturbation, the deformation of the Wigner function occurs more gradually than in the GHZ(3) state. This proves the distributed type entanglement and comparatively higher robustness of the W(3) state \cite{De2003Multiqubit}, even under averaged gaussian-distributed amplitude perturbation.
\begin{figure}
\centering
\begin{tabular}{ccccccc}
\xincludegraphics[width=0.9\linewidth, label={a)}]{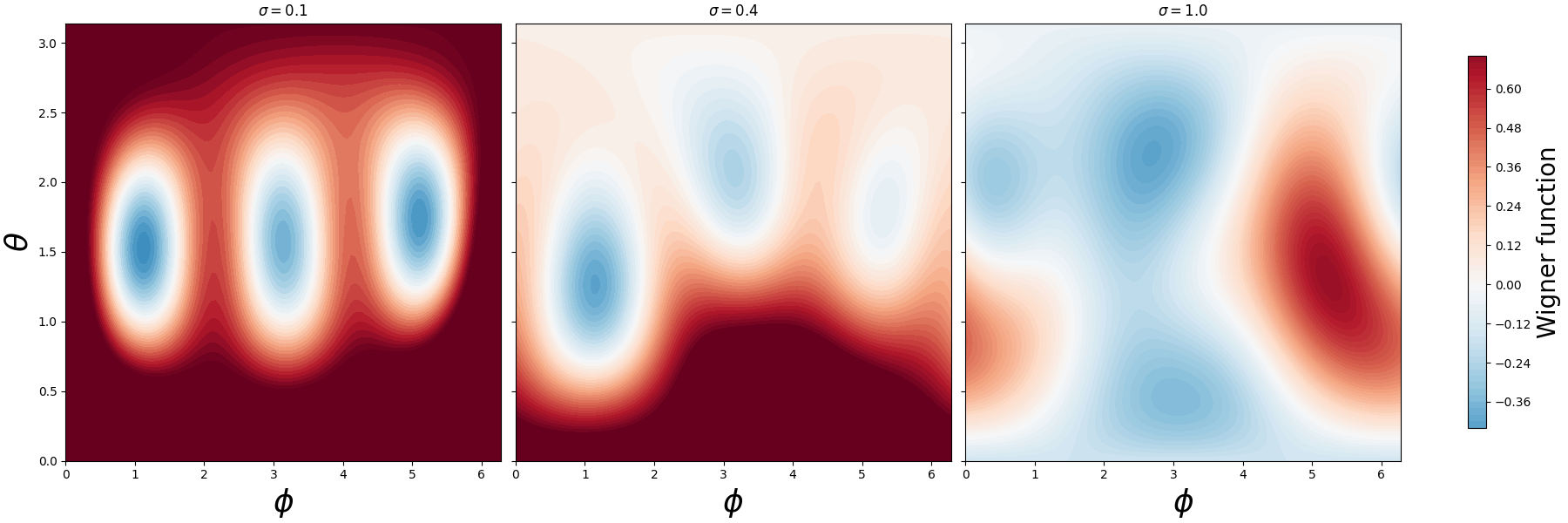} \\
\xincludegraphics[width=0.9\linewidth, label={b)}]{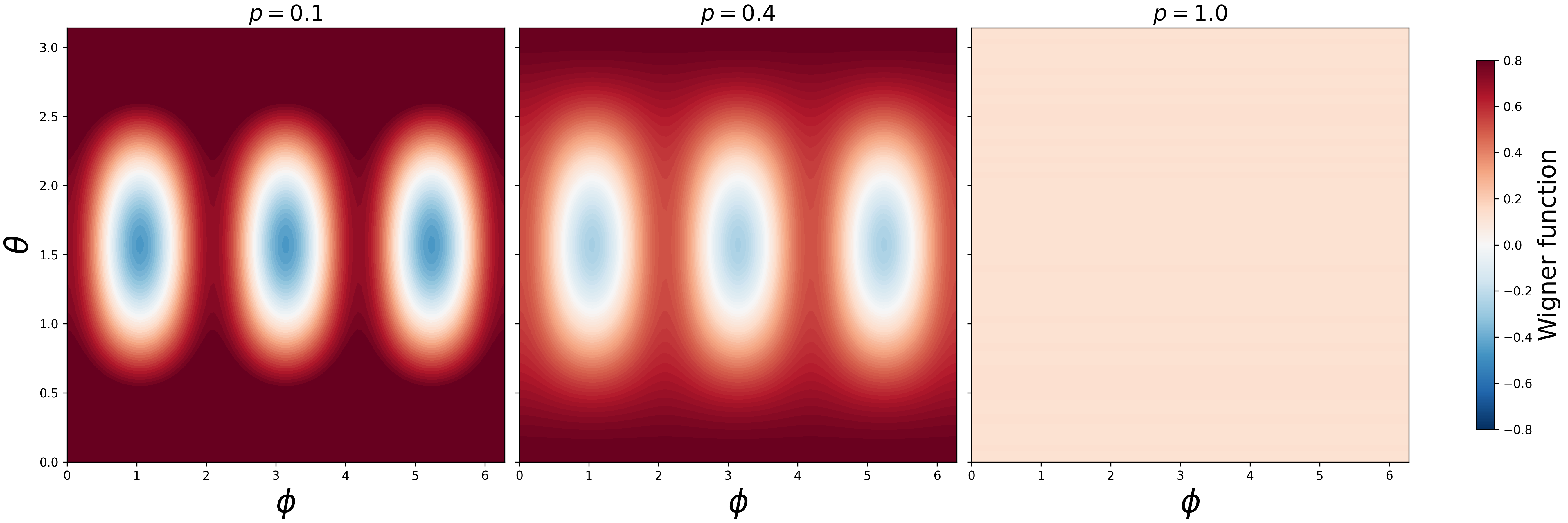} 
\end{tabular}
    \caption{Equal-angle spin Wigner function of GHZ(3) state (a) under gaussian-distributed amplitude perturbation with increasing standard deviation $\sigma = 0.1, 0.4, 1.0$ leading to a gradual deformation of phase space structure, (b) under white noise with increasing noise parameter $p=0.1, 0.4, 1.0$, moving the state rapidly to a uniform, classical-like distribution.}
\end{figure}
\begin{figure}
\centering
\begin{tabular}{ccccccc}
\xincludegraphics[width=0.9\linewidth, label={a)}]{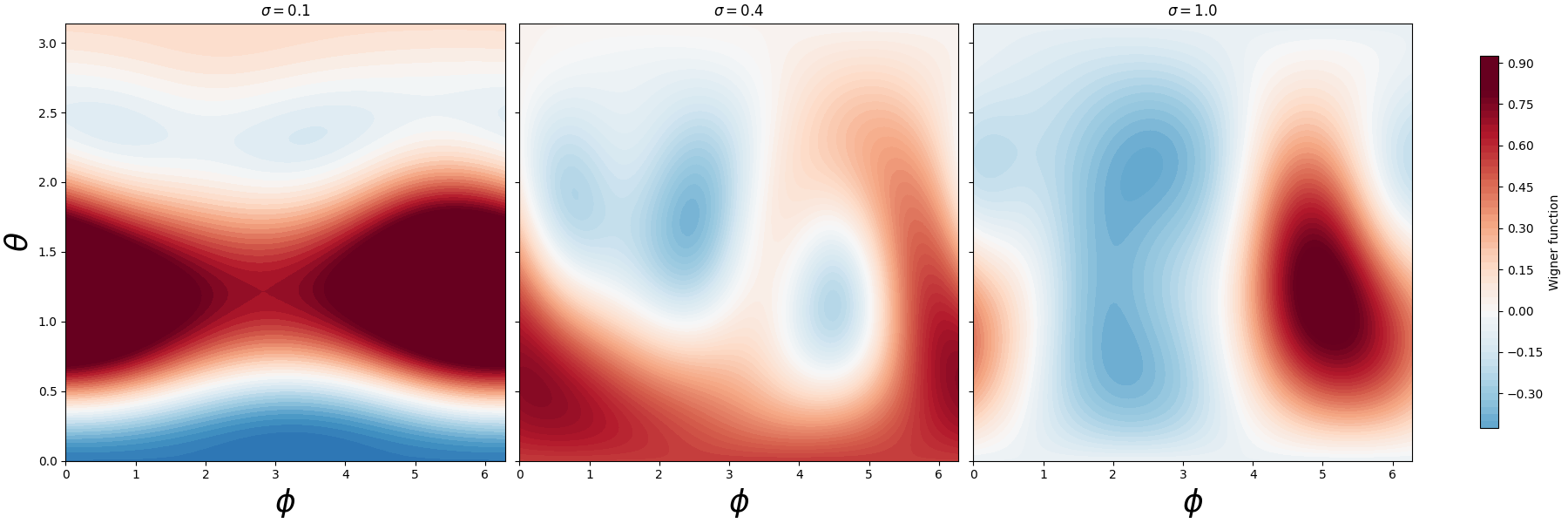} \\
\xincludegraphics[width=0.9\linewidth, label={b)}]{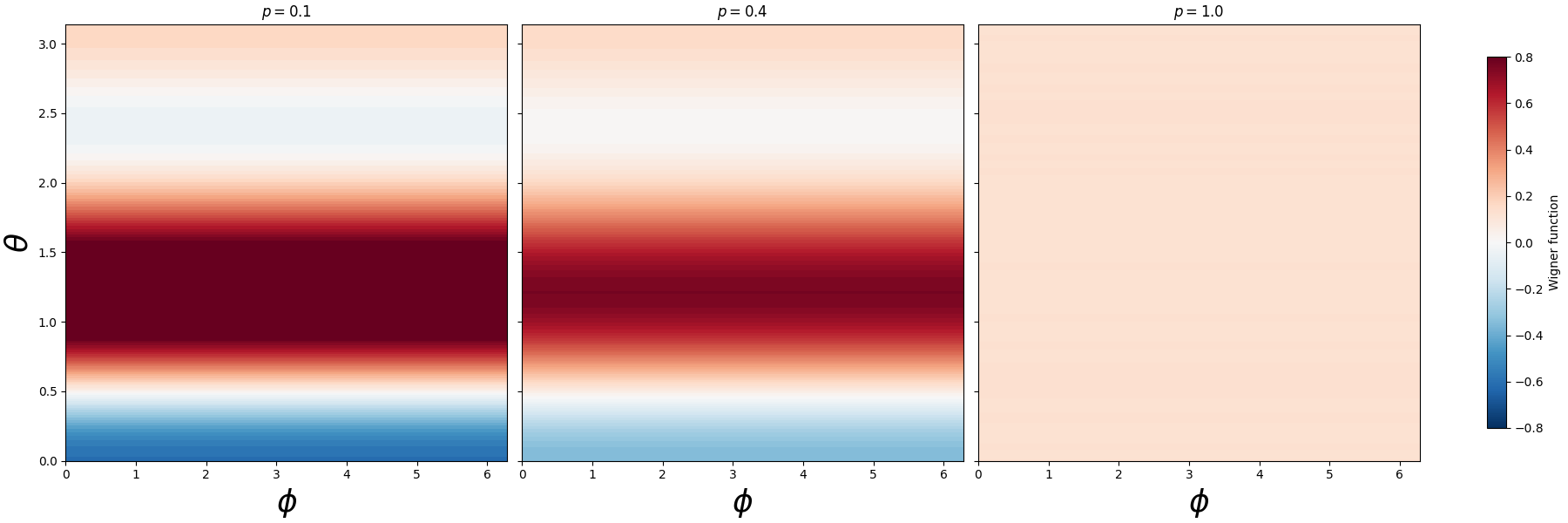}
\end{tabular}
    \caption{ Equal-angle spin Wigner function of W(3) state (a) under gaussian-distributed amplitude perturbation with increasing standard deviation $\sigma = 0.1, 0.4, 1.0$, leading to a gradual deformation of phase space structure, and (b) under white noise with increasing noise parameter $p=0.1, 0.4, 1.0$, moving the state rapidly to a uniform, classical-like distribution.}
\end{figure}
\begin{figure}
\centering
\begin{tabular}{ccccccc}
\xincludegraphics[width=0.9\linewidth, label={a)}]{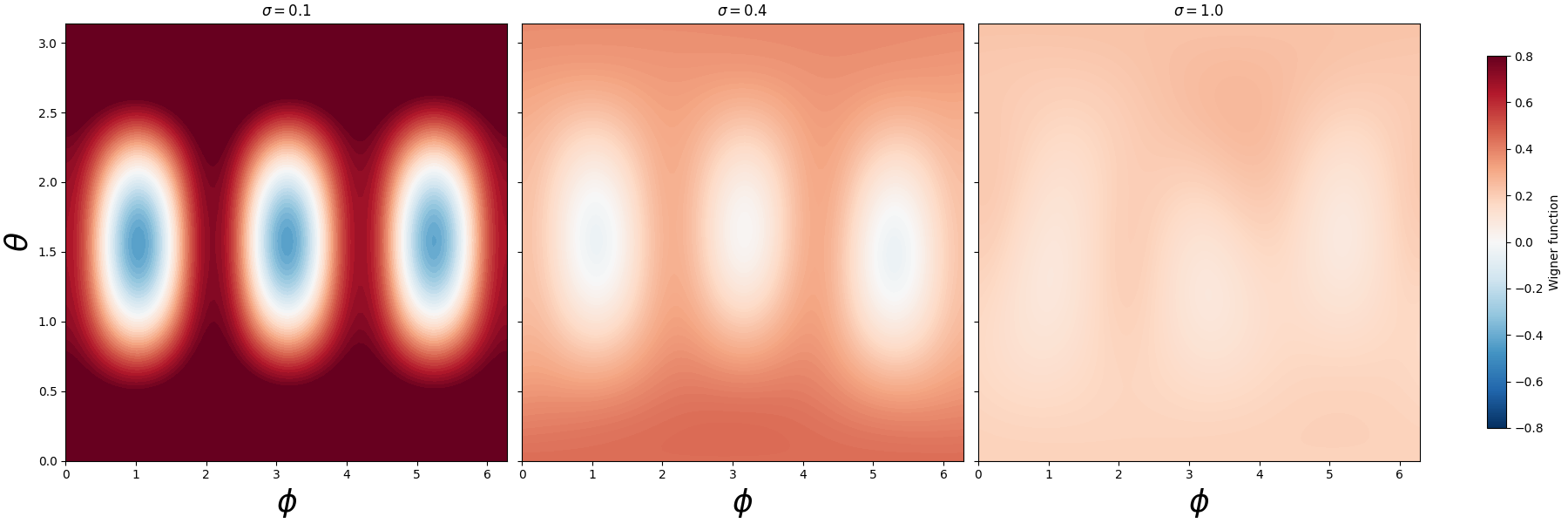} \\
\xincludegraphics[width=0.9\linewidth, label={b)}]{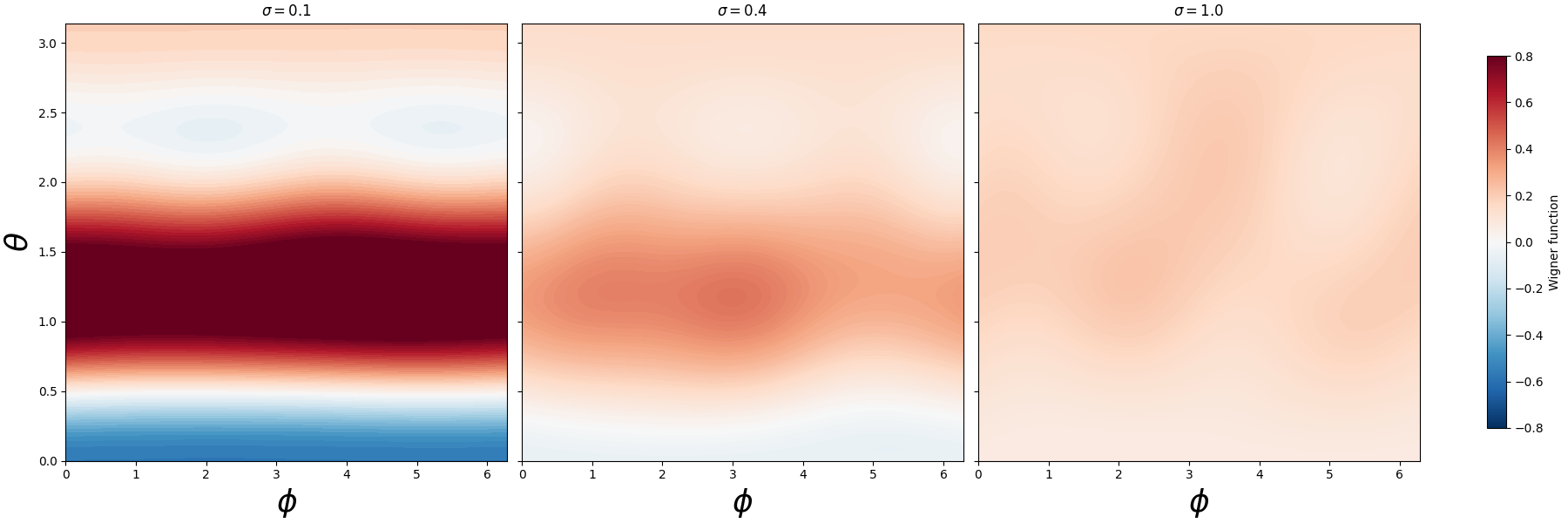} 

\end{tabular}
    \caption{The ensemble-averaged spin Wigner function for (a) the GHZ(3) state and (b) the W(3) state is presented under Gaussian-distributed amplitude perturbations with varying strengths of $\sigma = 0.1, 0.4, 1.0$.  The ensemble-averaged spin Wigner function suppresses realization-dependent fluctuations. }
\end{figure}
\section{Conclusion}
In this work, we have investigated the response of two fundamentally different types of tripartite entangled states, namely the GHZ(3) and W(3) states, under two types of classical noise models—Gaussian and white noise—using both qualitative and quantitative approaches. Starting from the ideal probability distribution, we emphasized how noise redistributes their probability among other computational basis states as noise increases. While both states undergo decoherence under increasing noise strength, their robustness differs depending on their respective type of entanglement.

Uhlmann-Jozsa fidelity was used as a quantitative measure to study the overall degradation of the states under noise. Our results show that, under both Gaussian and white noise, the fidelity curves of GHZ(3) and W(3) states nearly overlap across the entire noise range. This validates that fidelity serves as a quantity to capture the global loss of similarity with the ideal state. But, it fails to distinguish how different types of multipartite entangled states respond to noise based on their entanglement structure, consistent with known limitations of fidelity-based measures in noisy quantum systems.

To overcome this limitation, we employed the equal-angle spin Wigner function as the qualitative measure, which offers direct visualization of non-classical features in phase space. For ideal states, the Wigner function revealed distinct signatures of entanglement, such as a sharp interference pattern and strong negativity for the GHZ(3) state and a smooth, delocalized band-like structure for the W(3) state. Under gaussian-distributed amplitude perturbation, the single-realization Wigner function demonstrated gradual deformation of phase-space structure, capturing only the immediate impact of noise on the quantum features. In contrast, under white noise, rapid and uniform suppression of both positive and negative ranges was seen, driving the system toward a constant phase-space structure independent of ($\theta$, $\phi$).

The ensemble-averaged Wigner function reveals the effect of gaussian-distributed amplitude perturbation by eliminating realization-dependent fluctuations. Here, both the states are noted to evolve towards mixed-state phase-space distribution with increasing noise strength. The W(3) state was noted to exhibit a more gradual deformation compared to the GHZ(3) state. This confirmed its comparatively higher robustness against both noise types.

Overall, this study highlights that while fidelity provides a quantitative measure of decoherence, phase-space representations such as the spin Wigner function are essential because they uncover the qualitative differences in entanglement structure and robustness. Our combined analysis presented in this work offers a framework for understanding how tripartite GHZ(3) and W(3) entangled states degrade under noise models, and this may be extended to more complex entangled states and realistic noise models relevant to quantum information processing.

\section{COMPETING INTERESTS}
No, I declare that the authors have no competing interests
as defined by Springer, or other interests that might
be perceived to influence the results and/or discussion
reported in this paper.
\section{DATA AVAILABILITY}
No, I do not have any research data outside the submitted
manuscript file.
\section{AUTHOR CONTRIBUTIONS}
Nithya Priya: Conceptualization, methodology,
literature review, data curation, formal analysis, statistical
analysis, writing, Reviewing, Editing; S.Saravana Veni : Visualization, Investigation,
Validation, Writing, Reviewing and Editing. The
authors has approved the final manuscript.Araceli Venegas-Gomez : Investigation and Validation; Ria Rushin Joseph : Revieing and visualization.

\end{document}